\documentclass{PoS}

\title{Multiquark states in the covariant quark confinement model}

\ShortTitle{Multiquark states in the covariant quark confinement model}

\author{\speaker{Mikhail A. Ivanov}\\
        JINR\\
        E-mail: \email{ivanovm@theor.jinr.ru}}

\abstract{This talk reviews the last applications of the covariant quark model 
for studying  the properties of the multiquark states: 
$B_s-$meson (quark-antiquark state), light baryons (three-quark states)
and tetraquark (four-quark state).
The form factors of the $B(B_s)\to P(V)-$transitions are 
evaluated in the full kinematical region of momentum transfer squared.  
The widths of some $B_s-$nonleptonic decays are calculated. 
The static properties of the proton and neutron, and 
the $\Lambda$-hyperon (magnetic moments and charge radii) and the behavior 
of the nucleon form factors at low momentum transfers are described. 
The consequences of treating the X(3872) meson as a tetraquark
bound state are explored. The decay widths of the observed
channels $X\to J/\psi+2\pi (3\pi)$ and $X\to \bar D^0 +  D^0+\pi^0$ via the
intermediate off--shell states $X\to J/\psi+\rho(\omega)$ and 
$X\to \bar D +  D^{\,\ast}$ are  calculated. 
Its one-photon decay $X\to \gamma+ J/\psi $ is also analyzed. 
}

\FullConference{XXI International Baldin Seminar on High Energy 
Physics Problems,\\
		September 10-15, 2012\\
		JINR, Dubna, Russia}

\begin{document}

\section{Introduction}

The covariant quark model with infrared confinement
developed in a series of papers 
(see Refs.~\cite{Branz:2009cd}-\cite{Gutsche:2012ze}) 
is a successful tools for a unified description
of the multiquark states: mesons, baryons, tetraquarks, etc. 
The covariant quark model is an effective quantum
field approach to hadronic interactions based on the
interaction Lagrangian between hadrons and their
constituent quarks. Knowing a corresponding interpolating quark current 
allows calculating the matrix element of physical processes 
in a consistent way. A distinctive feature of this approach is that 
the multiquark states,  such as baryons (three quarks), tetraquarks 
(four quarks), etc.,  can be considered and
described as rigorously as the simplest quark-antiquark systems (mesons). 
The coupling constants between hadrons and their interpolating quark 
currents are determined from the connection condition
$Z_H = 0$ proposed in Refs.~\cite{Weinberg:1962hj,Salam:1962ap} 
and used further in numerous subfields of particle physics (for a review,
see Refs.~\cite{Hayashi:1967hk,Efimov:1988yd,Efimov:1993ei}). 
Here $Z_H$ is a renormalization constant of the hadron wave function. 
The matrix elements of physical processes are determined by a set of 
associated quark diagrams, which are constructed according
to $1/N_c-$expansion. In the covariant quark model an
infrared cutoff is effectively introduced in the space of
Fock--Schwinger parameters, which are integrated
out in the expressions for the matrix elements. Such a
procedure allows one to eliminate all the threshold
singularities associated with quark production and
thereby ensures quark confinement. The model has no
ultraviolet divergences due to vertex hadron--quark
form factors, which describe a nonlocal structure of
hadrons. The covariant quark model has a few free
parameters: a mass of constituent quarks, an infrared
cutoff parameter that characterizes confinement
region, and parameters that describe an effective size
of hadrons.

We review here the last applications of the covariant quark model 
for studying  the properties of the $B_s-$meson, the light baryons and
tetraquarks. The form factors of the $B(B_s)\to P(V)-$transitions are 
evaluated in the full kinematical region of momentum transfer squared.  
As an application of the obtained results 
the widths of the $B_s-$nonleptonic decays are calculated. 
The modes  
$D_s^- D_s^+,$  $D_s^{\,\ast\,-} D_s^{+}+D_s^- D_s^{\,\ast\,+}$ and
$D_s^{\,\ast\,-} D_s^{\,\ast\,+}$ give the largest contribution
to $\Delta\Gamma$ for the $B_s-\bar B_s$ system.
The mode $J/\psi\phi$ is suppressed by the color factor
but it is interesting for the search of CP-violating 
New-Physics possible effects in the $B_s-\bar B_s$ mixing.

The static properties of the proton and neutron, and 
the $\Lambda$-hyperon (magnetic moments and charge radii) and the behavior 
of the nucleon form factors at low momentum 
transfers are described. The conservation of gauge invariance of
the electromagnetic transition matrix elements in the presence of a nonlocal 
coupling of the baryons to the three constituent quark fields is discussed. 

The consequences of treating the X(3872) meson as a tetraquark
bound state are explored. 
The decay widths of the observed
channels $X\to J/\psi+2\pi (3\pi)$ and $X\to \bar D^0 +  D^0+\pi^0$ via the
intermediate off--shell states $X\to J/\psi+\rho(\omega)$ and 
$X\to \bar D +  D^{\,\ast}$ are  calculated. 
Its one-photon decay $X\to \gamma+ J/\psi $ is also analyzed. 
The matrix element of the transition
$X\to \gamma+ J/\psi $ is calculated and its gauge invariance is proved.
For reasonable values of the size parameter $\Lambda_X$ of the 
X(3872) consistency with the available experimental data is found.
The possible impact of the X(3872) in a s-channel dominance
description of the $J/\psi$ dissociation cross section is discussed.

\section{Covariant quark model}

The coupling of a hadron $H$ to its constituent
quarks is described  by the Lagrangian:

\begin{equation}
\label{eq:Lagr}
{\cal L}_{{\rm int}} = g_H\cdot H(x) \cdot J_H(x)
\end{equation}
where the quark currents are defined as

\begin{eqnarray*}
J_M(x) &=& \int\!\! dx_1 \!\!\int\!\! dx_2\,F_M (x,x_1,x_2)\cdot
\bar q^a_{f_1}(x_1)\, \Gamma_M \,q^a_{f_2}(x_2)
 \qquad {\rm Meson}
\\
&&\\
J_B(x) &=& \int\!\! dx_1 \!\!\int\!\! dx_2 \!\!\int\!\! dx_3\,
F_B (x,x_1,x_2,x_3) 
\\
&\times& \Gamma_1 \, q^{a_1}_{f_1}(x_1) \, 
\Big( q^{a_2}_{f_2}(x_2) C \, \Gamma_2 \, q^{a_3}_{f_3}(x_3)
\Big)\cdot \varepsilon^{a_1a_2a_3}   
 \qquad {\rm Baryon}
\\
&&\\
J_T(x) &=& 
\int\!\! dx_1\ldots\int\!\! dx_4\, 
F_T (x,x_1,\ldots,x_4) 
\\
&\times&  
\Big(
q_{f_1}^{a_1}(x_1)\, C\Gamma_1\, q_{f_2}^{a_2}(x_2)
\Big) \cdot
\Big(\bar q_{f_3}^{a_3}(x_3)\, \Gamma_2 C\, \bar q_{f_4}^{a_4}(x_4)\Big)
\cdot 
\varepsilon^{a_1a_2c} \varepsilon^{a_3a_4c}
\qquad {\rm Tetraquark}
\end{eqnarray*}
where $\Gamma$ is a Dirac matrix or a string of Dirac matrices
which projects onto the spin quantum number of the hadron $H(x)$.
The matrix $C=\gamma^{0}\gamma^{2}$ is
the usual charge conjugation matrix and the $a_i$ $(i=1,2,3)$ are color 
indices.
The function $F_H$ is related to the scalar part of the
Bethe-Salpeter amplitude and characterizes the finite size of the
hadron. To satisfy translational invariance the function $F_H$
has to fulfil the identity 
$F_H(x+a,x_1+a,\ldots,x_n+a)=F_H(x,x_1,\ldots,x_n)$ 
for any four-vector ``a''. In the following we use
a specific  form for the scalar vertex function
\begin{equation}
\label{eq:vertex}
F_H(x,x_1,\ldots,x_n)=\delta\left(x -\sum\limits_{i=1}^n w_i x_i \right) 
\Phi_H\Big(\sum\limits_{i<j}((x_i-x_j)^2\Big)\,,
\end{equation}
where $\Phi_H$ is the correlation function of the constituent quarks
with masses $m_{i}, (i=1,\ldots,n)$ and the mass ratios
$w_i = m_{i}/\sum\limits_{j=1}^n m_j$.

The coupling constant $g_H$ in Eq.~(\ref{eq:Lagr}) is determined by the
so-called {\it compositeness condition} originally proposed 
in Refs.~\cite{Weinberg:1962hj,Salam:1962ap} and extensively used in 
Refs.~\cite{Hayashi:1967hk,Efimov:1988yd,Efimov:1993ei}. 
The compositeness condition requires that the renormalization constant of
the elementary hadron field $H(x)$ is set to zero
\begin{equation}
\label{eq:Z eq 0}
Z_H \, = \, 1 - \, g^2_H\,\Pi'_H(m^2_H) \, = \, 0
\end{equation}
where $\Pi^\prime_H$ is the derivative of the hadron mass operator.
To clarify the physical meaning of the compositeness  condition 
in Eq.~(\ref{eq:Z eq 0}), we first want to remind the reader
that the renormalization constant $Z_H^{1/2}$ can also interpreted as
the matrix element between the physical and the corresponding bare state.
The condition  $Z_H=0$ implies that the physical state does not contain
the bare state and is appropriately described as a bound state.
The interaction Lagrangian of Eq.~(\ref{eq:Lagr}) and
the corresponding free parts of the Lagrangian describe
both the constituents (quarks) and the physical particles (hadrons)
which are viewed as the bound states of the quarks.
As a result of the interaction, the physical particle is dressed,
i.e. its mass and wave function have to be renormalized.
The condition $Z_H=0$ also effectively excludes
the constituent degrees of freedom from the space of physical states.
It thereby guarantees that there is no double counting
for the physical observable under consideration.
The constituents exist only in  virtual states.
One of the corollaries of the compositeness condition is the absence
of a direct interaction of the dressed charged particle with the
electromagnetic field. Taking into account both the tree-level
diagram and the diagrams with the self-energy insertions into the
external legs (i.e. the tree-level diagram times $Z_H -1$) yields
a common factor $Z_H$  which is equal to zero. 

We have used free fermion propagators for the quarks given by
\begin{equation}
\label{eq:quark-prop}
S_i(k)=\frac{1}{m_i-\not\! k}
\end{equation}
with an effective constituent quark mass $m_i$. 

For calculational convenience we will choose a simple Gaussian form for the 
vertex function $\bar \Phi_H(-\,k^2)$. 
The minus sign in the argument of this function is chosen to emphasize 
that we are working in Minkowski space. One has
\begin{equation}
\bar \Phi_H(-\,k^2) 
= \exp\left(k^2/\Lambda_H^2\right)
\label{eq:Gauss}
\end{equation}
where the parameter $\Lambda_H$ characterizes the size of the hadron $H$.
Since $k^2$ turns into $-\,k_E^2$ in Euclidean space the form
(\ref{eq:Gauss}) has the appropriate fall-off behavior in the Euclidean region.
We emphasize that any choice for  $\Phi_H$ is appropriate
as long as it falls off sufficiently fast in the ultraviolet region of
Euclidean space to render the corresponding Feynman diagrams ultraviolet 
finite. As mentioned before we shall choose a Gaussian form for $\Phi_H$
for calculational convenience.

We have included the confinement of quarks 
to our model in Ref.~\cite{Branz:2009cd}.
It was done, first, by introducing the scale integration
in the space of $\alpha$-parameters, and, second, by cutting this 
scale integration on the upper limit which corresponds to an infrared cutoff. 
In this manner one removes all possible thresholds 
presented in the initial quark diagram. 
The cutoff parameter is taken to be the same for all physical processes.
We have adjusted other model parameters by fitting the calculated
quantities of the basic physical processes to available experimental data.

Let us give the basic features of the infrared confinement in our model.
All physical matrix elements are described by the Feynman diagrams which
are the convolution of the free quark propagators and vertex functions. 
 Let $n$, $\ell$ and  $m$ be the number
of the propagators, loops and vertices, respectively.
In Minkowski space the $\ell$-loop diagram will be represented as
\begin{eqnarray}
&&
\Pi(p_1,...,p_m) = 
\int\!\! [d^4k]^\ell  
\prod\limits_{i_1=1}^{m} \,
\Phi_{i_1+n} \left( -K^2_{i_1+n}\right)
\prod\limits_{i_3=1}^n\, S_{i_3}(\tilde k_{i_3}+v_{i_3}),
\nonumber\\
&&\nonumber\\
&&
K^2_{i_1+n} =\sum_{i_2}(\tilde k^{(i_2)}_{i_1+n}+v^{(i_2)}_{i_1+n})^2
\label{eq:diag}
\end{eqnarray}
where the vectors $\tilde k_i$  are  linear combinations 
of the loop momenta $k_i$. The $v_i$ are  linear combinations 
of the external momenta $p_i$ to be specified in the following.
The strings of Dirac matrices appearing in the calculation need not concern 
us since they do not depend on the momenta. 
The external momenta $p_i$ are all chosen to be ingoing such that one has 
$\sum\limits_{i=1}^m p_i=0$. 
 
Using the Schwinger representation of the local quark propagator one has
\[
S(k) = (m+\not\! k)
\int\limits_0^\infty\! 
d\beta\,e^{-\beta\,(m^2-k^2)}\, \qquad (k^2<m^2)\,.
\]
For the vertex functions one takes the Gaussian form. One has
\begin{equation}
\label{eq:vert} 
\Phi_{i+n} \left( -K^2\right)\,
 =
\exp\left[\beta_{i+n}\,K^2\right] \qquad i=1,...,m\, ,
\end{equation}
where the parameters $\beta_{i+n}=s_{i}=1/\Lambda^2_{i}$ are 
related to the size parameters. The integrand in Eq.~(\ref{eq:diag})
has a Gaussian form with the exponential $kak+2kr+R$ where $a$ is
$\ell\times\ell$ matrix depending on the parameter $\beta_i$,
$r$ is the $\ell$-vector composed from the external momenta, and
$R$ is a quadratic form of the external momenta.
Tensor loop integrals are calculated with the help of the differential
representation 
\[
k_i^\mu e^{2kr} = \frac{1}{2}\frac{\partial}{\partial r_{i\,\mu}}e^{2kr},
\] 
We have written a FORM \cite{Vermaseren:2000nd} program that achieves 
the necessary commutations of the differential operators in a very efficient 
way.
 After doing the loop integrations one obtains
\[
\Pi =  \int\limits_0^\infty d^n \beta \, F(\beta_1,\ldots,\beta_n) \,,
\]
where $F$ stands for the whole structure of a given diagram. 
The set of Schwinger parameters $\beta_i$ can be turned into a simplex by 
introducing an additional $t$--integration via the identity 
\[ 
1 = \int\limits_0^\infty dt \, \delta(t - \sum\limits_{i=1}^n \beta_i)
\] 
leading to 
\begin{equation}
\hspace*{-0.2cm}
\Pi   = \int\limits_0^\infty\! dt t^{n-1}\!\! \int\limits_0^1\! d^n \alpha \, 
\delta\Big(1 - \sum\limits_{i=1}^n \alpha_i \Big) \, 
F(t\alpha_1,\ldots,t\alpha_n). 
\label{eq:loop_2} 
\end{equation}
There are altogether $n$ numerical integrations: $(n-1)$ $\alpha$--parameter
integrations and the integration over the scale parameter $t$. 
The very large $t$-region corresponds to the region where the singularities
of the diagram with its local quark propagators start appearing. 
However, as described in \cite{Branz:2009cd}, if one introduces 
an infrared cut-off on the upper limit of the t-integration, all 
singularities vanish because the integral is now convergent for any value
of the set of kinematic variables.
We cut off the upper integration at $1/\lambda^2$ and obtain
\[
  \Pi^c = \!\!  
\int\limits_0^{1/\lambda^2}\!\! dt t^{n-1}\!\! \int\limits_0^1\! d^n \alpha \, 
\delta\Big(1 - \sum\limits_{i=1}^n \alpha_i \Big) \, 
F(t\alpha_1,\ldots,t\alpha_n).
\]  
By introducing the infrared cut-off one has removed all potential thresholds 
in the quark loop diagram, i.e. the quarks are never on-shell and are thus
effectively confined. We take the cut-off parameter $\lambda$ to be the 
same in all physical processes.
The numerical evaluations have been done by a numerical program 
written in the FORTRAN code.

As a further illustration of the infrared confinement effect relevant to the 
applications in this paper 
we consider the case of a scalar one--loop two--point 
function. One has
\[
\Pi_2(p^2) =
\!\!\int\! \frac{d^4k_E}{\pi^2} 
\frac{e^{-sk_E^2}}{[m^2 + (k_E+\frac12 p_E)^2]
                  [m^2 + (k_E-\frac12 p_E)^2]} 
\]
where we have collected all the nonlocal 
Gaussian vertex form factors in the numerator factor $e^{-sk_E^2}$. Note that
the momenta $k_E$, $p_E$ are Euclidean momenta. Doing the 
loop integration one obtains 
\begin{eqnarray}
\Pi_2(p^2) &=&\!\! \int\limits_0^\infty\!\! dt \frac{t}{(s+t)^2} 
\int\limits_0^1\!\! d\alpha \, 
\exp\Big[ - t z_{\,\rm loc}+ \frac{st}{s+t} z_1 \Big],
\nonumber\\
&&\nonumber\\
z_{\,\rm loc} &=& m^2 - \alpha(1-\alpha)p^2,
\qquad
z_1 =  \Big(\alpha - \frac{1}{2}\Big)^2 p^2.  
\end{eqnarray} 
The integral $\Pi_2(p^2)$ can be seen to have a branch point at $p^2=4m^2$
because $z_{\rm loc}$ is zero when $\alpha=1/2$. 
By introducing a  cut-off  on the $t$-integration one obtains  
\begin{equation} 
\Pi^c_2(p^2) =\!\! \int\limits_0^{1/\lambda^2}\!\! dt \frac{t}{(s+t)^2} 
\int\limits_0^1\! d\alpha \, 
\exp\Big[ - t z_{\,\rm loc}+ \frac{st}{s+t} z_1\Big]\,.
\label{eq:conf}
\end{equation} 
The one-loop two-point function $\Pi_2^c(p^2)$ Eq.(\ref{eq:conf}) can be seen
to have no branch point at $p^2=4m^2$. 

The gauging of the nonlocal Lagrangian in Eq.~(\ref{eq:Lagr}) proceeds 
in a way suggested in 
Refs.~\cite{Mandelstam:1962mi,Terning:1991yt}
and used before by us (see, for instance,
Refs.~~\cite{Ivanov:1996pz,Faessler:2006ft}).
In order to guarantee local invariance of
the nonlocal Lagrangian in Eq.~(\ref{eq:Lagr}) one multiplies 
each quark field $q(x_i)$ with a gauge field exponential:

\begin{equation}
q_i(x_i)\to e^{-ie_{q_1} I(x_i,x,P)} \, q_i(x_i)\, 
\label{eq:gauging}
\end{equation} 
where
\begin{equation}
I(x_i,x,P) = \int\limits_x^{x_i} dz_\mu A^\mu(z). 
\label{eq:path}
\end{equation} 
The path $P$ connects the end-points of the path integral.
One then expands the gauge exponential
up to the requisite power of $e_{q}A_\mu$ needed in the perturbative series. 
We need to know only the derivatives of the path integral
expressions when calculating the perturbative series.
Therefore, we use the 
formalism suggested in~\cite{Mandelstam:1962mi,Terning:1991yt} 
which is based on the path-independent definition of the derivative of 
$I(x,y,P)$: 
\begin{equation} 
\lim\limits_{dx^\mu \to 0} dx^\mu 
\frac{\partial}{\partial x^\mu} I(x,y,P) \, = \, 
\lim\limits_{dx^\mu \to 0} [ I(x + dx,y,P^\prime) - I(x,y,P) ]
\label{eq:path1}
\end{equation}  
where the path $P^\prime$ is obtained from $P$ by shifting the end-point $x$
by $dx$.
The definition (\ref{eq:path1}) leads to the key rule
\begin{equation} 
\frac{\partial}{\partial x^\mu} I(x,y,P) = A_\mu(x)
\label{path2}
\end{equation} 
which in turn states that the derivative of the path integral $I(x,y,P)$ does 
not depend on the path $P$ originally used in the definition. 

As a result of this rule we are getting the part of the Lagrangian 
which describes the nonlocal interaction
of the hadron, quark and electromagnetic fields to the first order in the 
electromagnetic charge.

\section{$B_s$-meson}

We give below the necessary definitions of the leptonic decay constants,
invariant form factors and helicity amplitudes.

The leptonic decay constants of the pseudoscalar
and vector mesons are defined by
\begin{eqnarray}
\label{eq:lept}
N_c\, g_P\! \int\!\! \frac{d^4k}{ (2\pi)^4 i}\, \widetilde\Phi_P(-k^2)\,
{\rm tr} \biggl[O^{\,\mu} S_1(k+w_1 p) \gamma^5 S_2(k-w_2 p) \biggr] 
&=&f_P\, p^\mu, \phantom{m_V}\qquad p^2=m^2_P,
\nonumber\\
N_c\, g_V\! \int\!\! \frac{d^4k}{ (2\pi)^4 i}\, \widetilde\Phi_V(-k^2)\,
{\rm tr} \biggl[O^{\,\mu} S_1(k+w_1 p)\not\!\varepsilon_V  S_2(k-w_2 p) \biggr] 
&=& m_V f_V\, \varepsilon_V^\mu,\qquad p^2=m^2_V,
\end{eqnarray}
where $N_c=3$ is the number of colors.

Herein our primary subjects are the following matrix elements, 
which can be expressed  via dimensionless form factors:


\begin{eqnarray}
&&
\langle 
P^{\,\prime}_{[\bar q_1 q_3]}(p_2)\,|\,\bar q_2\, O^{\,\mu}\, q_1\,| P_{[\bar q_3 q_2]}(p_1)
\rangle
\,=\,
\nonumber\\
&=&
N_c\, g_P\,g_{P^{\,'}}\!\!  \int\!\! \frac{d^4k}{ (2\pi)^4 i}\, 
\widetilde\Phi_P\Big(-(k + w_{13} p_1)^2\Big)\,
\widetilde\Phi_{P^{\,'}}\Big(-(k + w_{23} p_2)^2\Big)
\nonumber\\
&\times&
{\rm tr} \biggl[
O^{\,\mu}\, S_1(k+p_1)\, \gamma^5\, S_3(k)\, \gamma^5\, S_2(k+p_2) 
\biggr]
 \, = \, F_+(q^2)\, P^{\,\mu} + F_-(q^2)\, q^{\,\mu}\,,
\label{eq:PP'}
\end{eqnarray}

\begin{eqnarray}
&&
\langle 
P^{\,\prime}_{[\bar q_1 q_3]}(p_2)\,
|\,\bar q_2\, (\sigma^{\,\mu\nu}q_\nu) \, q_1\,| 
P_{[\bar q_3 q_2]}(p_1)
\rangle
\,=\,
\nonumber\\
&=&
N_c\, g_P\,g_{P^{\,'}} \!\! \int\!\! \frac{d^4k}{ (2\pi)^4 i}\, 
\widetilde\Phi_P\Big(-(k + w_{13} p_1)^2\Big)\,
\widetilde\Phi_{P^{\,'}}\Big(-(k + w_{23} p_2)^2\Big)
\nonumber\\
&\times&
{\rm tr} \biggl[ 
\sigma^{\,\mu\nu}q_\nu \,
S_1(k+p_1)\, \gamma^5\, S_3(k)\, \gamma^5 \,S_2(k+p_2) 
\biggr]
 \, = \,
\frac{i}{m_1+m_2}\,\left(q^2\, P^{\,\mu}-q\cdot P\, q^{\,\mu}\right)\,F_T(q^2),
\label{eq:PP'T}
\end{eqnarray}

\begin{eqnarray}
&&
\langle 
V(p_2,\varepsilon_2)_{[\bar q_1 q_3]}\,
|\,\bar q_2\, O^{\,\mu}\,q_1\, |\,P_{[\bar q_3 q_2]}(p_1)
\rangle
\,=\,
\nonumber\\
&=&
N_c\, g_P\,g_V \!\! \int\!\! \frac{d^4k}{ (2\pi)^4 i}\, 
\widetilde\Phi_P\Big(-(k + w_{13} p_1)^2\Big)\,
\widetilde\Phi_V\Big(-(k + w_{23} p_2)^2\Big) 
\nonumber\\
&\times&
{\rm tr} \biggl[ 
O^{\,\mu} \,S_1(k+p_1)\,\gamma^5\, S_3(k) \not\!\varepsilon_2^{\,\,\dagger} \,
S_2(k+p_2)\, \biggr]
\label{eq:PV}\\
 & = &
\frac{\varepsilon^{\,\dagger}_{\,\nu}}{m_1+m_2}\,
\left( - g^{\mu\nu}\,P\cdot q\,A_0(q^2) + P^{\,\mu}\,P^{\,\nu}\,A_+(q^2)
       + q^{\,\mu}\,P^{\,\nu}\,A_-(q^2) 
+ i\,\epsilon^{\mu\nu\alpha\beta}\,P_\alpha\,q_\beta\,V(q^2)\right),
\nonumber
\end{eqnarray}

\begin{eqnarray}
&&
\langle 
V(p_2,\varepsilon_2)_{[\bar q_1 q_3]}\,
|\,\bar q_2\, (\sigma^{\,\mu\nu}q_\nu(1+\gamma^5))\,q_1\, |\,P_{[\bar q_3 q_2]}(p_1)
\rangle 
\,=\,
\nonumber\\
 & = &
N_c\, g_P\,g_V \!\! \int\!\! \frac{d^4k}{ (2\pi)^4 i}\, 
\widetilde\Phi_P\Big(-(k + w_{13} p_1)^2\Big)\,
\widetilde\Phi_V\Big(-(k + w_{23} p_2)^2\Big)
\nonumber\\
& \times &
{\rm tr} \biggl[ 
(\sigma^{\,\mu\nu}q_\nu(1+\gamma^5))
\,S_1(k+p_1)\,\gamma^5\, S_3(k) \not\!\varepsilon_2^{\,\,\dagger} \,S_2(k+p_2)\, 
\biggr]
\label{eq:PVT}\\
 & = &
\varepsilon^{\,\dagger}_{\,\nu}\,
\left( - (g^{\mu\nu}-q^{\,\mu}q^{\,\nu}/q^2)\,P\cdot q\,a_0(q^2) 
       + (P^{\,\mu}\,P^{\,\nu}-q^{\,\mu}\,P^{\,\nu}\,P\cdot q/q^2)\,a_+(q^2)
+ i\,\epsilon^{\mu\nu\alpha\beta}\,P_\alpha\,q_\beta\,g(q^2)\right).
\nonumber
\end{eqnarray}
Here, $P=p_1+p_2$, $q=p_1-p_2$, $\varepsilon_2^\dagger\cdot p_2=0$,
$p_i^2=m_i^2$. Since there are three sorts of quarks involved in these
processes, we introduce the notation with two subscripts
$w_{ij}=m_{q_j}/(m_{q_i}+m_{q_j})$ $(i,j=1,2,3)$ so that $w_{ij}+w_{ji}=1$. 
The form factors defined in Eq.\,(\ref{eq:PVT}) satisfy the physical 
requirement $a_0(0)=a_+(0)$, which ensures that no kinematic singularity 
appears in the matrix element at $q^2=0$.  
For reference it is useful to relate the form factors we have defined 
to those used, e.g., in Ref.\,\cite{Khodjamirian:2006st}, 
which are denoted by a superscript $c$ in the following formulae:
\begin{eqnarray}
F_+ &=& f_+^{c}\,,\quad 
F_- = -\,\frac{m_1^2-m_2^2}{q^2}\,(f_+^{c} - f_0^{c})\,, \quad 
F_T = f_T^{c}\,, 
\nonumber\\
&&\nonumber\\
A_0 &=& \frac{m_1 + m_2}{m_1 - m_2}\,A_1^{c}\,, \quad 
A_+ = A_2^{c}\,,\quad
A_- =  \frac{2m_2(m_1+m_2)}{q^2}\,(A_3^{c} - A_0^{c})\,, \quad
V = V^{c}\,, 
\nonumber\\
&&\nonumber\\
a_0 &=& T_2^{c}\,, \quad g = T_1^{c}\,, \quad
a_+  =  T_2^{c}+\frac{q^2}{m_1^2-m_2^2}\,T_3^{c}\,.
\end{eqnarray}
We note in addition that the form factors $A_i^c(q^2)$ satisfy
the constraints: $A_0^{c}(0)=A_3^{c}(0)$ and 
\[
2m_2A_3^{c}(q^2) = (m_1+m_2) A_1^{c}(q^2) -(m_1-m_2) A_2^{c}(q^2)\,.
\]

It is convenient to express all physical observables
through the helicity form factors $H_m$.
The helicity form factors $H_m$ can be expressed in terms of
the invariant form factors in the following way 
(see Refs.~\cite{Ivanov:2000aj,Ivanov:2005fd,Ivanov:2006ni}):

\vspace{0.5cm}
\noindent
(a) Spin $S=0$:

\begin{eqnarray}
H_t &=& \frac{1}{\sqrt{q^2}}
\left\{(m_1^2-m_2^2)\, F_+ + q^2\, F_- \right\}\,,
\nonumber\\
H_\pm &=& 0\,,
\label{helS0b}\\
H_0 &=& \frac{2\,m_1\,|{\bf p_2}|}{\sqrt{q^2}} \,F_+ \,.
\nonumber
\end{eqnarray}

\vspace{0.5cm}
\noindent
(b) Spin $S=1$:

\begin{eqnarray}
H_t &=&
\frac{1}{m_1+m_2}\frac{m_1\,|{\bf p_2}|}{m_2\sqrt{q^2}}
\left\{ (m_1^2-m_2^2)\,(A_+ - A_0)+q^2 A_- \right\},
\nonumber\\
H_\pm &=&
\frac{1}{m_1+m_2}\left\{- (m_1^2-m_2^2)\, A_0
\pm 2\,m_1\,|{\bf p_2}|\, V \right\},
\label{helS1c}\\
H_0 &=&
\frac{1}{m_1+m_2}\frac{1}{2\,m_2\sqrt{q^2}}
\left\{-(m_1^2-m_2^2) \,(m_1^2-m_2^2-q^2)\, A_0
+4\,m_1^2\,|{\bf p_2}|^2\, A_+\right\},
\nonumber
\end{eqnarray}
where $|{\bf p_2}|=\lambda^{1/2}(m_1^2,m_2^2,q^2)/(2\,m_1)$
is the momentum  of the outgoing particles
in the rest frame of ingoing particle.

The effective Hamiltonian describing the $B_s$-nonleptonic
decays is given by (see, Ref.~\cite{Buchalla:1995vs})
\begin{eqnarray}
{\mathcal H}_{\rm eff} &=&
-\frac{G_F}{\sqrt{2}}\,V_{cb} V^\dagger_{cs}\,\sum_{i=1}^6 C_i\,Q_i,
\nonumber\\
&&\nonumber\\
Q_1 &=& (\bar c_{a_1} b_{a_2})_{V-A} (\bar s_{a_2} c_{a_1})_{V-A}, 
\qquad 
Q_2  =  (\bar c_{a_1}\,b_{a_1})_{V-A}, (\bar s_{a_2}\,c_{a_2})_{V-A},
\nonumber\\
Q_3 &=& (\bar s_{a_1} b_{a_1})_{V-A} (\bar c_{a_2} c_{a_2})_{V-A}, 
\qquad 
Q_4  =  (\bar s_{a_1} b_{a_2})_{V-A} (\bar c_{a_2} c_{a_1})_{V-A},
\nonumber\\
Q_5 &=& (\bar s_{a_1} b_{a_1})_{V-A} (\bar c_{a_2} c_{a_2})_{V+A}, 
\qquad 
Q_4  =  (\bar s_{a_1} b_{a_2})_{V-A} (\bar c_{a_2} c_{a_1})_{V+A},
\label{eq:Hamilt}
\end{eqnarray}
where the subscript $V-A$ refers to the usual left--chiral current
$O^\mu=\gamma^\mu(1-\gamma^5)$ and  $V+A$  to the usual right--chiral one
$O^\mu_+=\gamma^\mu(1+\gamma^5)$. The  $a_i$ denote the color indices. 

We consider the nonleptonic decays of the $B_s$-meson into
$ D_s^-\, D_s^+ $, 
$ D_s^- \, D_s^{\,\ast\,+} $, 
$ D_s^{\,\ast\,-} \,  D_s^+ $,
$ D_s^{\,\ast\,-} \, D_s^{\, \ast\,+} $ and
$ J/\psi\,\phi$. The calculation of the matrix elements is straightforward.
It directly leads to the representation corresponding to {\it naive} 
factorization. 

The widths can be conveniently expressed in terms of the helicity form factors
and leptonic decay constants.
In the case of the color-allowed decays one has
\begin{eqnarray}
\Gamma(B_s\to D_s^- D_s^+) &=&
\frac{G_F^2}{16\pi}\frac{|{\bf q_2}|}{m^2_{B_s}} [\lambda^{(s)}_c]^2 
\left( 
     C^{\,\rm eff}_{2}\,m_{D_s}\,f_{D_s}\,H_t^{B_sD_s}(m^2_{D_s})
 +2\,C^{\,\rm eff}_{6}\,f_{D_s}^{PS}\,F_S^{B_sD_s}(m^2_{D_s})
\right)^2\,,
\nonumber\\
\Gamma(B_s\to D_s^- D_s^{\,\ast\,+}) &=&
\frac{G_F^2}{16\pi}\frac{|{\bf q_{\,2}}|}{m^2_{B_s}} [\lambda^{(s)}_c]^2 
\left(     C^{\,\rm eff}_{2}\,m_{D_s}\,f_{D_s}\,H_t^{ B_s D^{\,\ast}_s}(m^2_{D_s})
       +2\,C^{\,\rm eff}_{6}\,\frac{m_{B_s}|{\bf q_{\,2}}|}{m_{D_s^{\,\ast}} }
  f_{D_s}^{PS}\,F_{PS}^{B_sD_s^\ast}(m^2_{D_s})
\right)^2\,,
\nonumber\\
\Gamma(B_s\to D_s^{\,\ast\,-} D_s^{+}) &=&
\frac{G_F^2}{16\pi}\frac{|{\bf q_{\,2}}|}{m^2_{B_s}} [\lambda^{(s)}_c]^2  
\left(     C^{\,\rm eff}_{2}\,m_{D^{\,\ast}_s}\,
f_{D^{\,\ast}_s}\,H_0^{B_sD_s}(m^2_{D^{\,\ast}_s})
\right)^2\,,
\nonumber\\
\Gamma(B_s\to D_s^{\,\ast\,-} D_s^{\,\ast\,+}) &=&
\frac{G_F^2}{16\pi}\frac{|{\bf q_{\,2}}|}{m^2_{B_s}} [\lambda^{(s)}_c]^2  
\left(     C^{\,\rm eff}_{2}\,m_{D^{\,\ast}_s}\,f_{D^{\,\ast}_s} \right)^2\,
\sum_{i=0,\pm}\left(H_i^{B_sD^{\,\ast}_s}(m^2_{D^{\,\ast}_s})\right)^2.
\label{eq:BsDD}
\end{eqnarray}
Here, $ \lambda^{(s)}_c=|V_{cb}V^\dagger_{cs}|$ and
$|{\bf q_2}|$ is the momentum  of the second outgoing particle
in the rest frame of $B_s-$meson.
The Wilson coefficients are combined as
$ C^{\,\rm eff}_{2}=C_2+\xi\, C_1+C_4+\xi\, C_3 $ and
$ C^{\,\rm eff}_{6}=C_6+\xi\, C_5.$ where a color factor $\xi=1/N_c$
will be suppressed in the numerical calculations according to 
$1/N_c-$expansion. Also we do not take into account the annihilation 
channels which are available for  the color-allowed decays.

The width of the color-suppressed $B_s\to J/\psi\,\phi$ decay 
is written as
\begin{eqnarray}
\Gamma(B_s\to J/\psi\,\phi ) &=&
\frac{G_F}{16\pi}\frac{|{\bf q_{\,2}}|}{m^2_{B_s}}  [\lambda^{(s)}_c]^2 
\left(C^{\,\rm eff}_1+ C^{\,\rm eff}_5\right)^2 
\left( m_{J/\psi}\,f_{J/\psi} \right)^2\,
\sum_{i=0,\pm}\left(H_i^{B_s J/\psi}(m^2_{J/\psi})\right)^2,
\label{eq:BsJpsiPhi}
\end{eqnarray}
where the  Wilson coefficients are combined as
$ C^{\,\rm eff}_{1}=C_1+\xi\, C_2+C_3+\xi\, C_4 $ and
$ C^{\,\rm eff}_{5}=C_5+\xi\, C_6.$.

The first application of our relativistic quark model with infrared confinement
to the description of the physical observables was done in our paper
\cite{Branz:2009cd}.  We have fitted the model parameters to
the leptonic and radiative decay constants of both pseudoscalar and vector
mesons. Then we have calculated transition form factors and 
the widths of the Dalitz decays and compared the results with
available experimental data. Here we calculate the form factors
describing the transitions of the heavy $B(B_s)-$mesons into
the light ones, e.g. $\pi,K,\rho,K^\ast,\phi.$ These quantities
are of great interest due to their applications to semileptonic,
nonleptonic and rare decays of the $B$ and $B_s-$mesons. 
Basically, they are calculated within the light-cone sum rules (LCSR)
in the region of large recoils (small transfer momentum squared).
Our approach allows one to evaluate the form factors in the full
kinematical regions including zero recoil.
First, we update the model parameters by fitting them to the leptonic
decay constants, see Table~\ref{tab:leptonic}, and
the widths of the radiative decays, see, Table~\ref{tab:em-widths}.
The results of the fit for the values of quark masses, the infrared cutoff
and the size parameters are given in Eqs.~(\ref{eq: fitmas}),
(\ref{eq:fitsize1}) and (\ref{eq:fitsize2}), respectively.

\begin{equation}
\def\arraystretch{2}
\begin{array}{cccccc}
     m_u        &      m_s        &      m_c       &     m_b & \lambda  &   
\\\hline
 \ \ 0.235\ \   &  \ \ 0.424\ \   &  \ \ 2.16\ \   &  \ \ 5.09\ \   & 
\ \ 0.181\ \   & \ {\rm GeV} 
\end{array}
\label{eq: fitmas}
\end{equation}

\begin{equation}
\def\arraystretch{2}
\begin{array}{ccccccccc}
 \Lambda_\pi   & \Lambda_K   & \Lambda_D        &  \Lambda_{D_s} &
 \Lambda_{B} & \Lambda_{B_s} & \Lambda_{B_c} &  \Lambda_{\rho}    & 
\\\hline
\ \ 0.87 \ \  & \ \ 1.04 \ \ & \ \ 1.47\ \  & \ \ 1.57 \ \ &
\ \ 1.88 \ \  & \ \ 1.95 \ \ & \ \ 2.42\ \  & \ \ 0.61 \ \ & \  {\rm GeV}
\end{array}
\label{eq:fitsize1}
\end{equation}

\begin{equation}
\def\arraystretch{2}
\begin{array}{ccccccccc}
 \Lambda_\omega   & \Lambda_\phi   & \Lambda_{J/\psi} &  \Lambda_{K^\ast} &
 \Lambda_{D^\ast}  & \Lambda_{D^\ast_s} & \Lambda_{B^\ast} &  \Lambda_{B^\ast_s}    & 
\\\hline
\ \ 0.47 \ \  & \ \ 0.88 \ \ & \ \ 1.48\ \ & \ \ 0.72 \ \ &
\ \ 1.16 \ \  & \ \ 1.17 \ \ & \ \ 1.72\ \ & \ \ 1.71 \ \ & \  {\rm GeV}
\end{array}
\label{eq:fitsize2}
\end{equation}

\begin{table}[t]
\caption{Leptonic decay constants $f_H$ (MeV) used in the least-squares
fit for our model parameters.}
\label{tab:leptonic}
\begin{center}
\def\arraystretch{1.5}
\begin{tabular}{cccc|}
\hline\hline
    & This work  & Other &  Ref.  \\
\hline
$f_\pi$  & 128.7 & $130.4\,\pm\, 0.2 $   & \cite{PDG,Rosner:2010ak}\\
%
%
$f_K$   & 156.1 & $156.1 \,\pm\, 0.8 $  & \cite{PDG,Rosner:2010ak}\\
%
%
$f_{D}$  & 205.9 & $206.7 \,\pm\, 8.9 $ & \cite{PDG,Rosner:2010ak}\\
%
%
$f_{D_s}$ & 257.5 & $257.5 \,\pm\, 6.1 $ & \cite{PDG,Rosner:2010ak}\\
%
%
$f_{B}$ & 191.1 & $192.8 \,\pm\, 9.9  $ & \cite{Laiho:2009eu}\\
%
%
$f_{B_s}$ & 234.9 & $238.8 \,\pm\, 9.5 $ & \cite{Laiho:2009eu}\\
%
%
$f_{B_c}$ & 489.0 & $489 \,\pm\, 5 $ & \cite{Chiu:2007km}\\
%
%
$f_{\rho}$ & 221.1 & $221 \,\pm\, 1 $ & \cite{PDG}\\
\hline\hline
\end{tabular}
\begin{tabular}{|cccc}
\hline\hline
    & This work  &  Other &  Ref.  \\
\hline
$f_\omega$  & 198.5  & $198 \,\pm\, 2 $ & \cite{PDG} \\
%
%
$f_\phi$    & 228.2  & $227\,\pm\,  2 $ & \cite{PDG} \\
%
%
$f_{J/\psi}$  & 415.0  & $415\,\pm\, 7  $ & \cite{PDG} \\
%
%
$f_{K^\ast}$  & 213.7  & $217\,\pm\, 7  $ & \cite{PDG} \\
%
%
$f_{D^\ast}$   & 243.3  & $245\,\pm\, 20  $ & \cite{Becirevic:1998ua} \\
%
%
$f_{D^\ast_s}$   & 272.0  & $272\,\pm\,26  $ & \cite{Becirevic:1998ua} \\
%
%
$f_{B^\ast}$   & 196.0  & $196\,\pm\, 44 $ & \cite{Becirevic:1998ua} \\
%
$f_{B_s^\ast}$   & 229.0  & $229\,\pm\, 46  $ & \cite{Becirevic:1998ua} \\
\hline\hline
\end{tabular}
\end{center}
\end{table}

\begin{table}[ht]
\begin{center}
\def\arraystretch{1.5}
\caption{Electromagnetic decay widths (keV) used in the least-squares fit 
for our model parameters.}
\label{tab:em-widths}
\vspace*{0.2cm}
\begin{tabular}{lcc}
\hline\hline 
Process & This work & Data~\cite{PDG}  \\
\hline
$\pi^0\to\gamma\gamma$          & \,\,  $ 5.06 \times 10^{-3}$ \,\,& 
                                \,\,$(7.7 \pm 0.4) \times 10^{-3}$\,\,\\ 
$\eta_c\to\gamma\gamma$         & 1.61 & 1.8 $\pm$ 0.8 \\ 
$\rho^{\pm}\to\pi^{\pm}\gamma$    & 76.0     &  67 $\pm$ 7  \\
$\omega\to\pi^0\gamma$          & 672      &  703 $\pm$ 25    \\
$K^{\ast \pm}\to K^\pm\gamma$      & 55.1     &  50 $\pm$ 5          \\
$K^{\ast 0}\to K^0\gamma$         & 116      &  116 $\pm$ 10      \\
$D^{\ast \pm}\to D^\pm\gamma$      & 1.22     &  1.5 $\pm$ 0.5 \\
$J/\psi \to \eta_c \gamma $      & 1.43     &  1.58 $\pm$ 0.37  \\
\hline
\end{tabular}
\end{center}
\end{table}

\begin{figure}[ht]
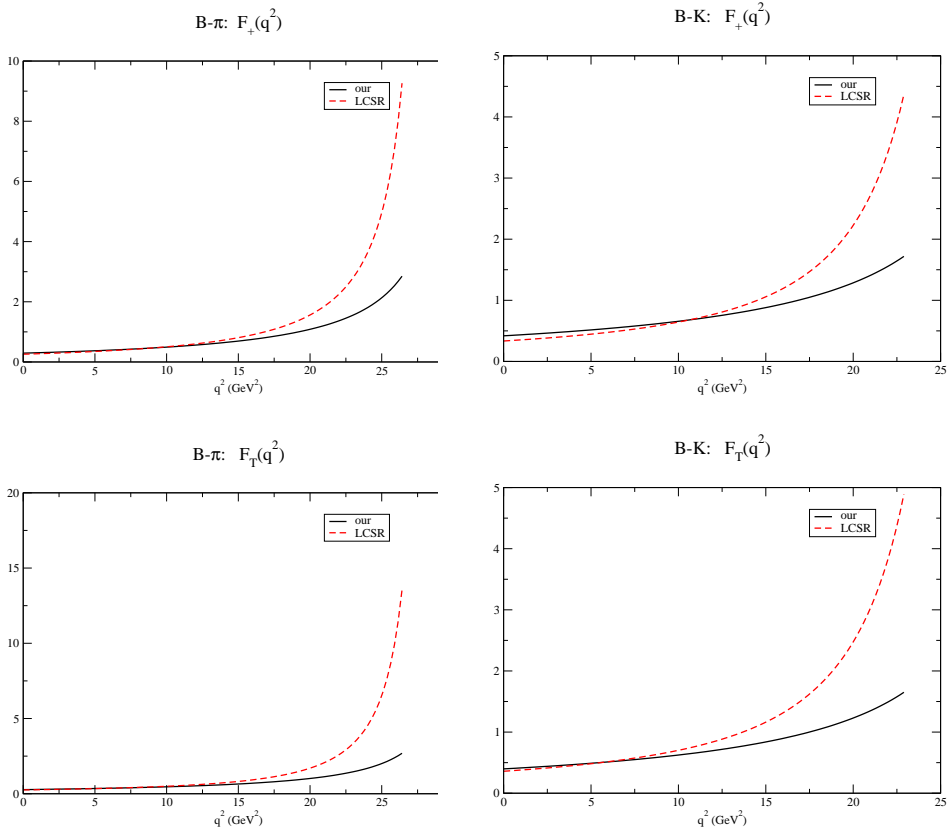

\begin{center}
\hspace*{-0.5cm}
\begin{tabular}{lr}
\includegraphics[width=0.40\textwidth]{Bpi_Fp.eps} &
\includegraphics[width=0.40\textwidth]{BK_Fp.eps} \\[2ex]
\includegraphics[width=0.40\textwidth]{Bpi_FT.eps} &
\includegraphics[width=0.40\textwidth]{BK_FT.eps}
\end{tabular}
\end{center}
\caption{\label{fig:ff-BP}
Our results for the form factors appearing 
in Eqs.\,(\protect\ref{eq:PP'}) 
      \& (\protect\ref{eq:PP'T}) -- \emph{Left panel},  
$B-\pi-$transition; 
                                and \emph{right panel}, 
$B-K-$transition. 
For comporison we plot the curves given by LCSR
from Ref.~\protect\cite{Ball}.
}
\end{figure}

\begin{figure}[ht]
\begin{center}
\hspace*{-0.5cm}
\begin{tabular}{lr}
\includegraphics[width=0.40\textwidth]{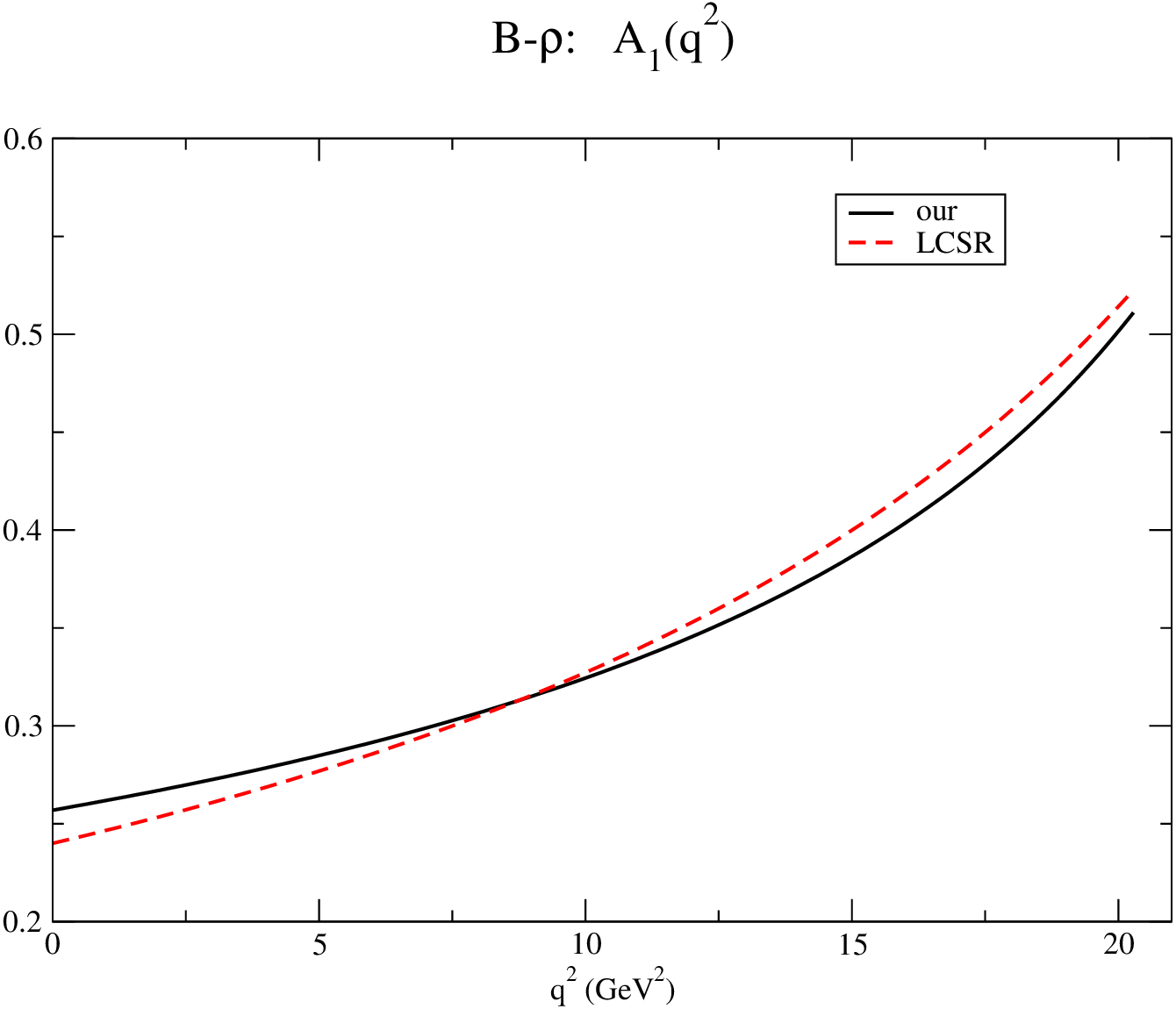} &
\includegraphics[width=0.40\textwidth]{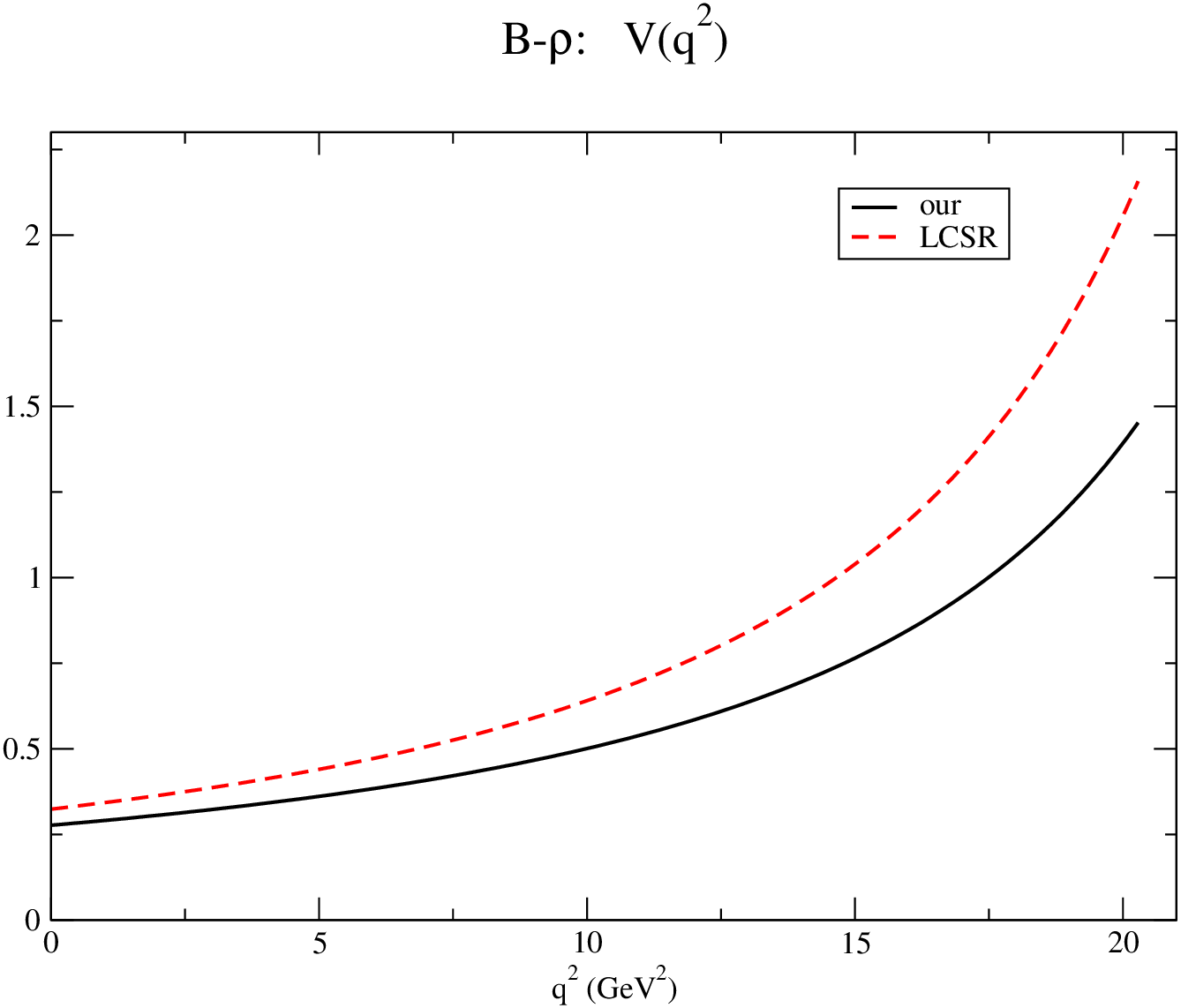} \\[2ex]
\includegraphics[width=0.40\textwidth]{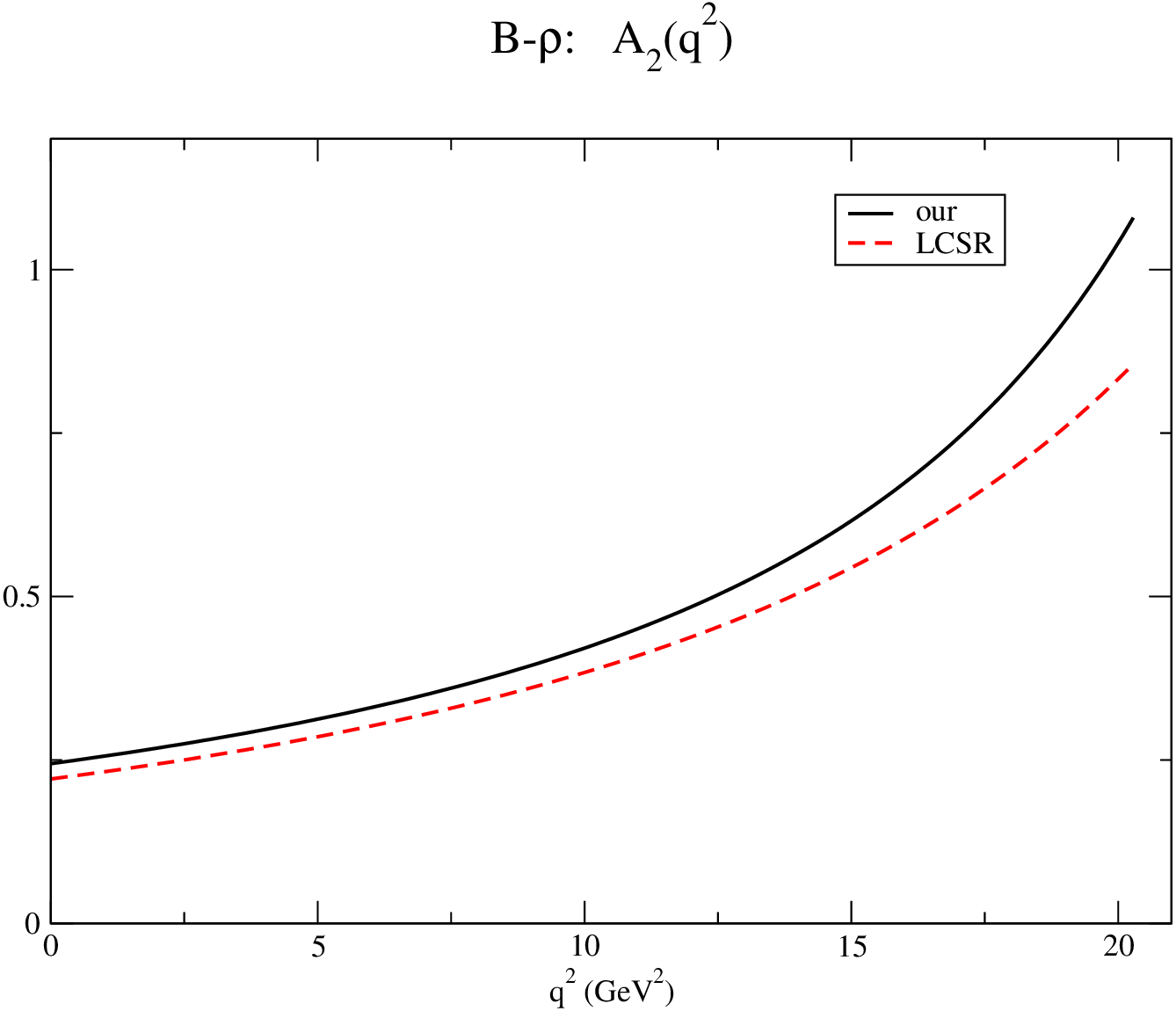} &
\includegraphics[width=0.40\textwidth]{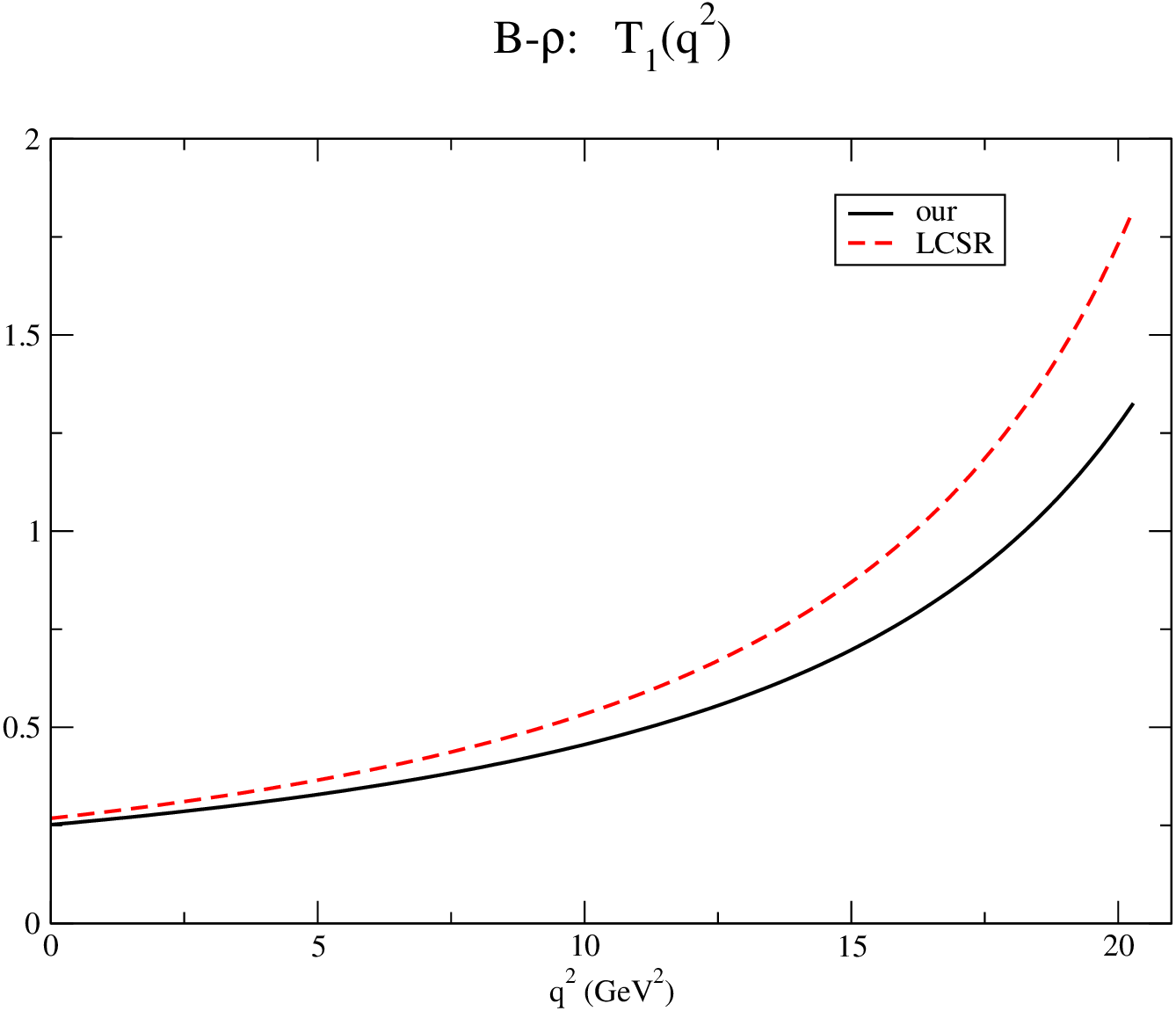}
\end{tabular}
\end{center}
\caption{\label{fig:ff-Brho}
Our results for the form factors appearing 
in Eqs.\,(\protect\ref{eq:PV}) 
      \& (\protect\ref{eq:PVT}) for  $B-\rho-$transition. 
For comporison we plot the curves given by LCSR
from Ref.~\protect\cite{Ball}. }
\end{figure}

\begin{figure}[ht]
\begin{center}
\hspace*{-0.5cm}
\begin{tabular}{lr}
\includegraphics[width=0.40\textwidth]{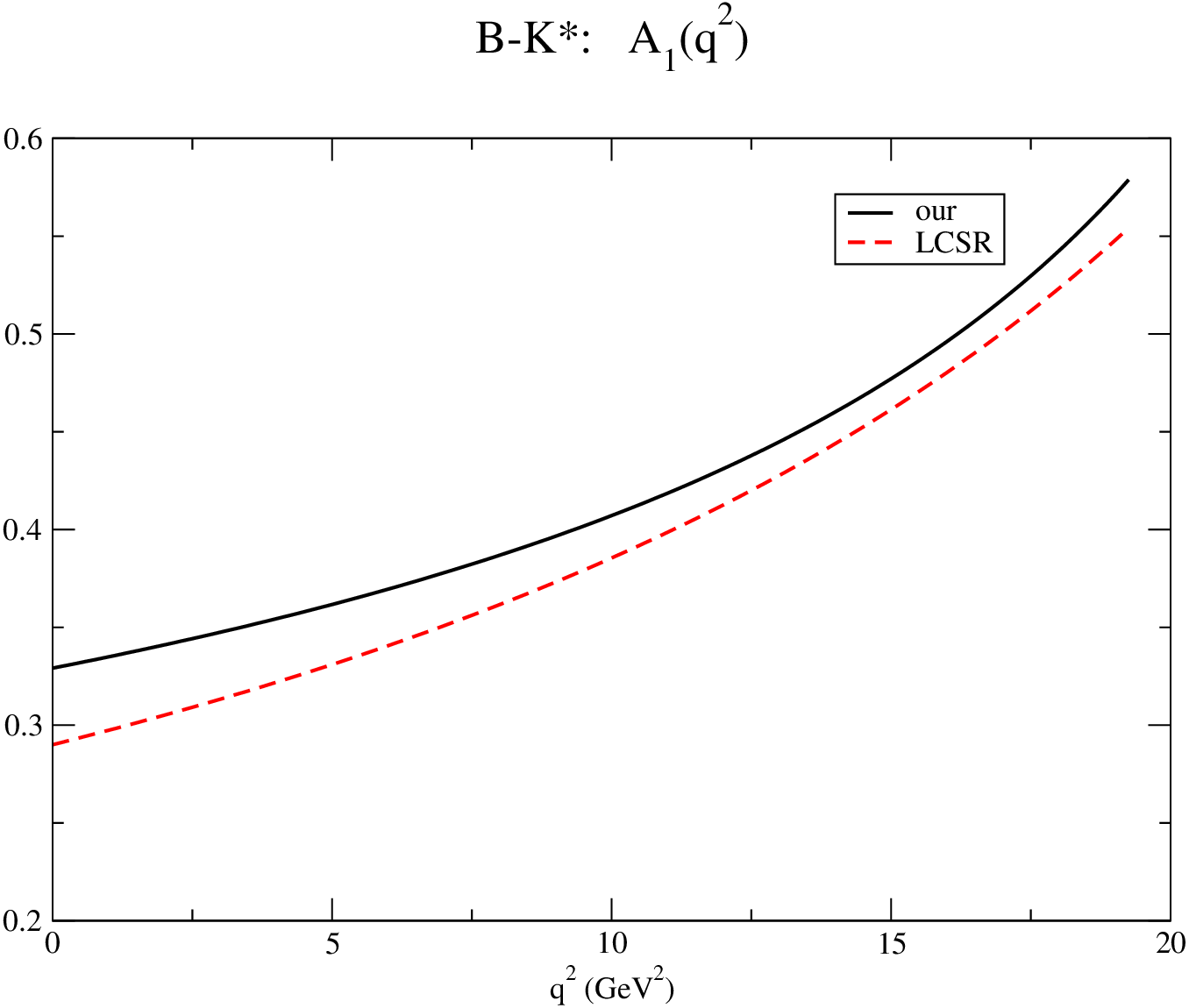} &
\includegraphics[width=0.40\textwidth]{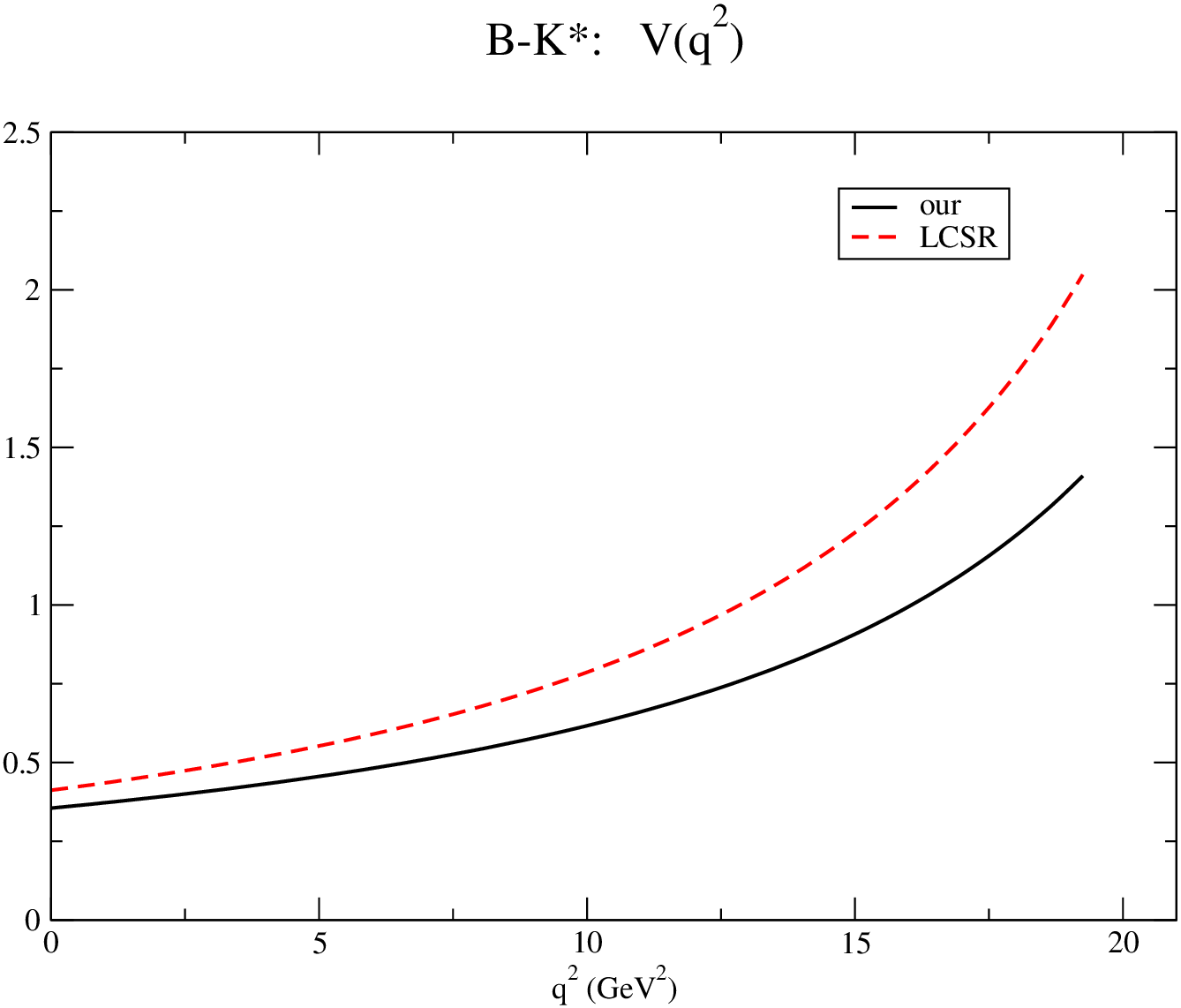} \\[2ex]
\includegraphics[width=0.40\textwidth]{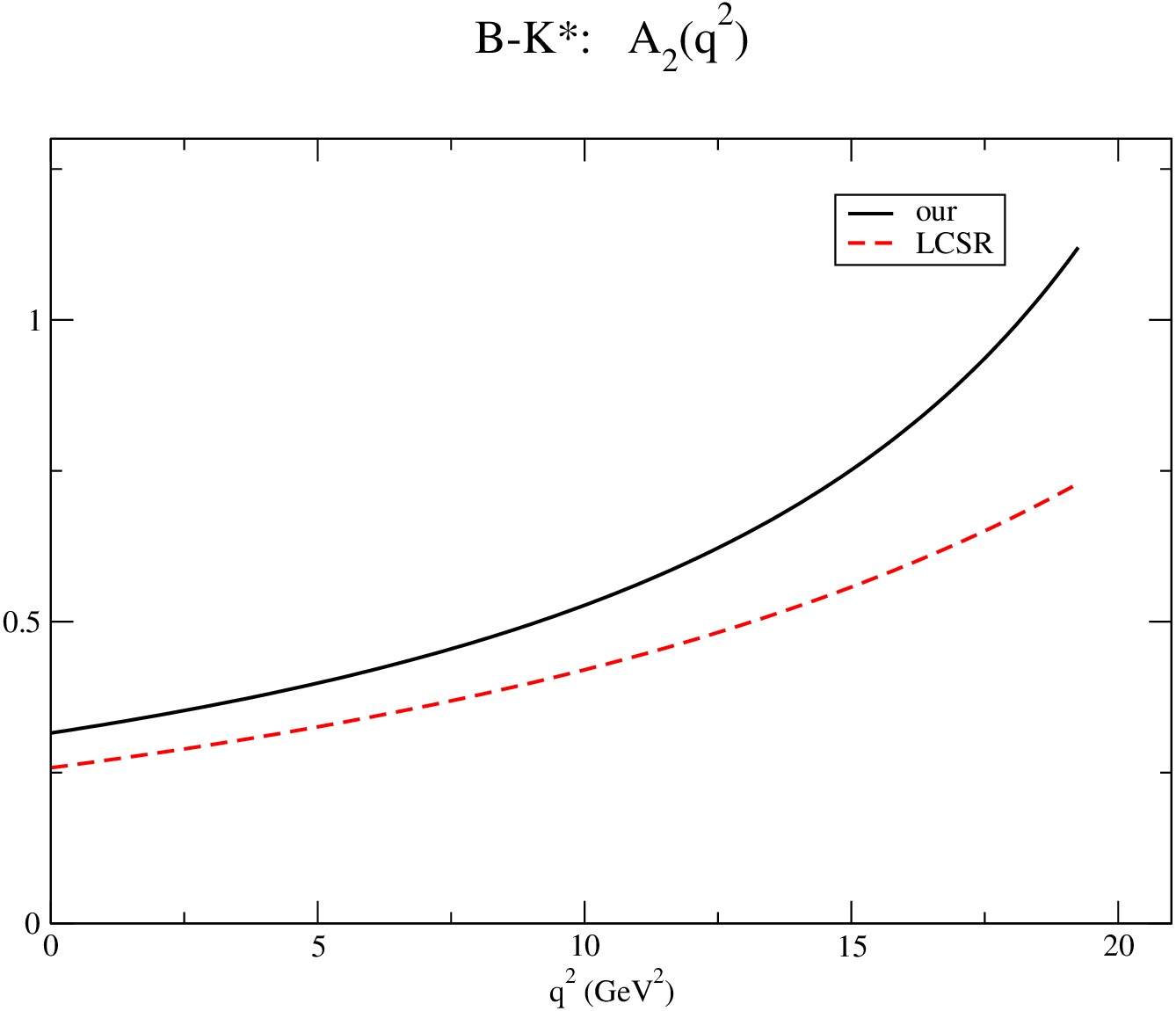} &
\includegraphics[width=0.40\textwidth]{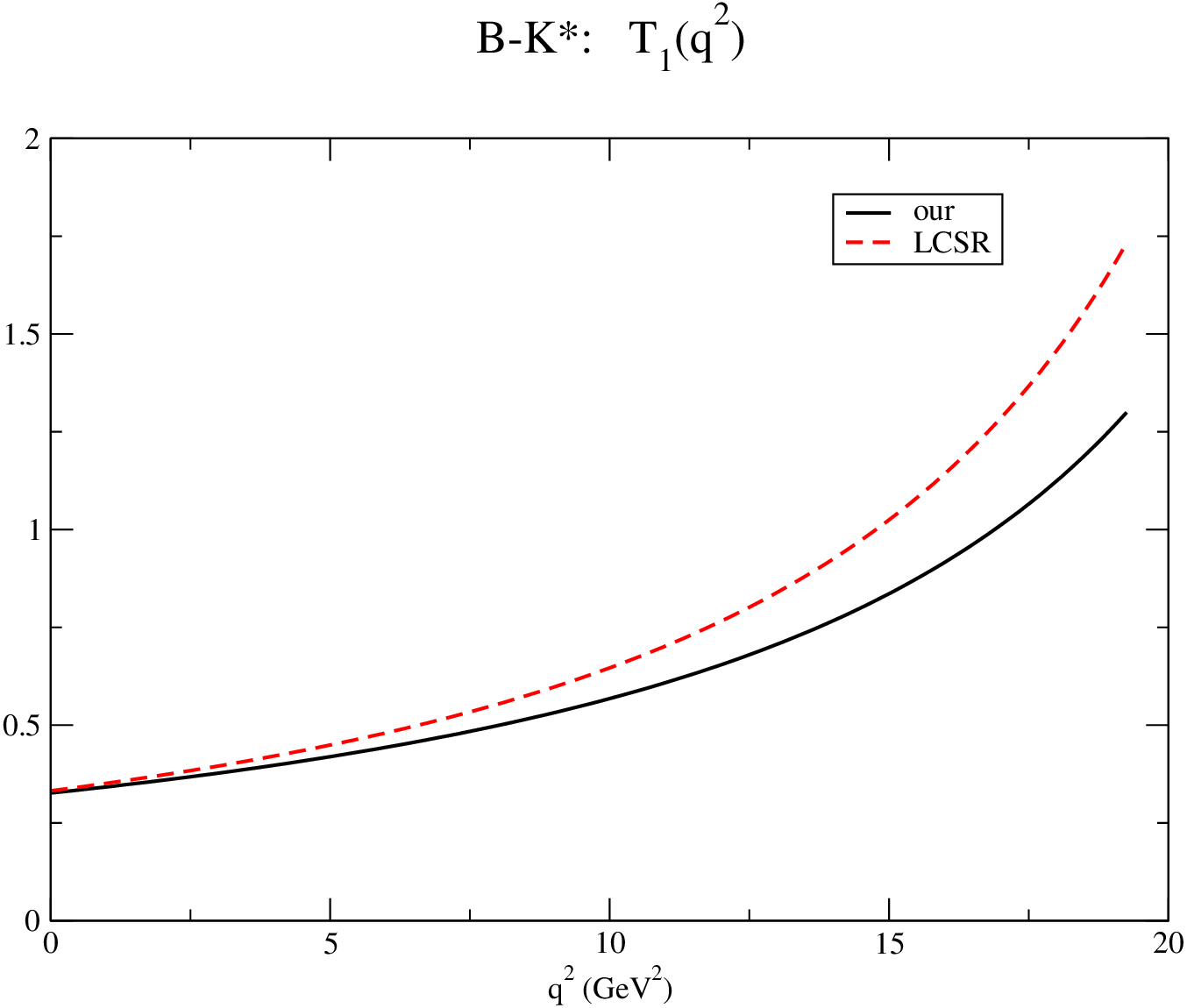}
\end{tabular}
\end{center}
\caption{\label{fig:ff-BKv}
Our results for the form factors appearing 
in Eqs.\,(\protect\ref{eq:PV}) 
      \& (\protect\ref{eq:PVT}) for  $B-K^\ast-$transition. 
For comporison we plot the curves given by LCSR
from Ref.~\protect\cite{Ball}. }
\end{figure}

\begin{figure}[ht]
\begin{center}
\hspace*{-0.5cm}
\begin{tabular}{lr}
\includegraphics[width=0.40\textwidth]{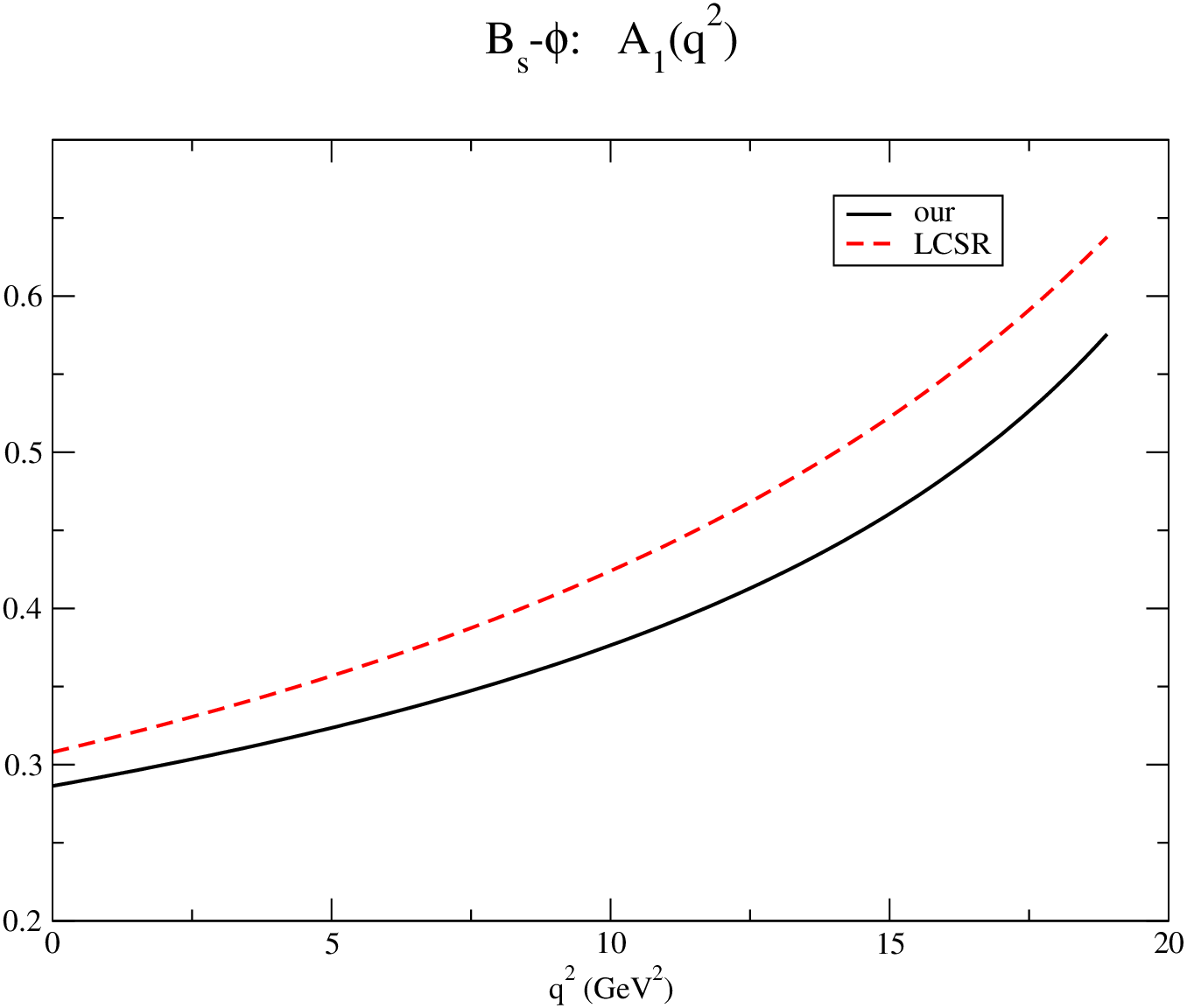} &
\includegraphics[width=0.40\textwidth]{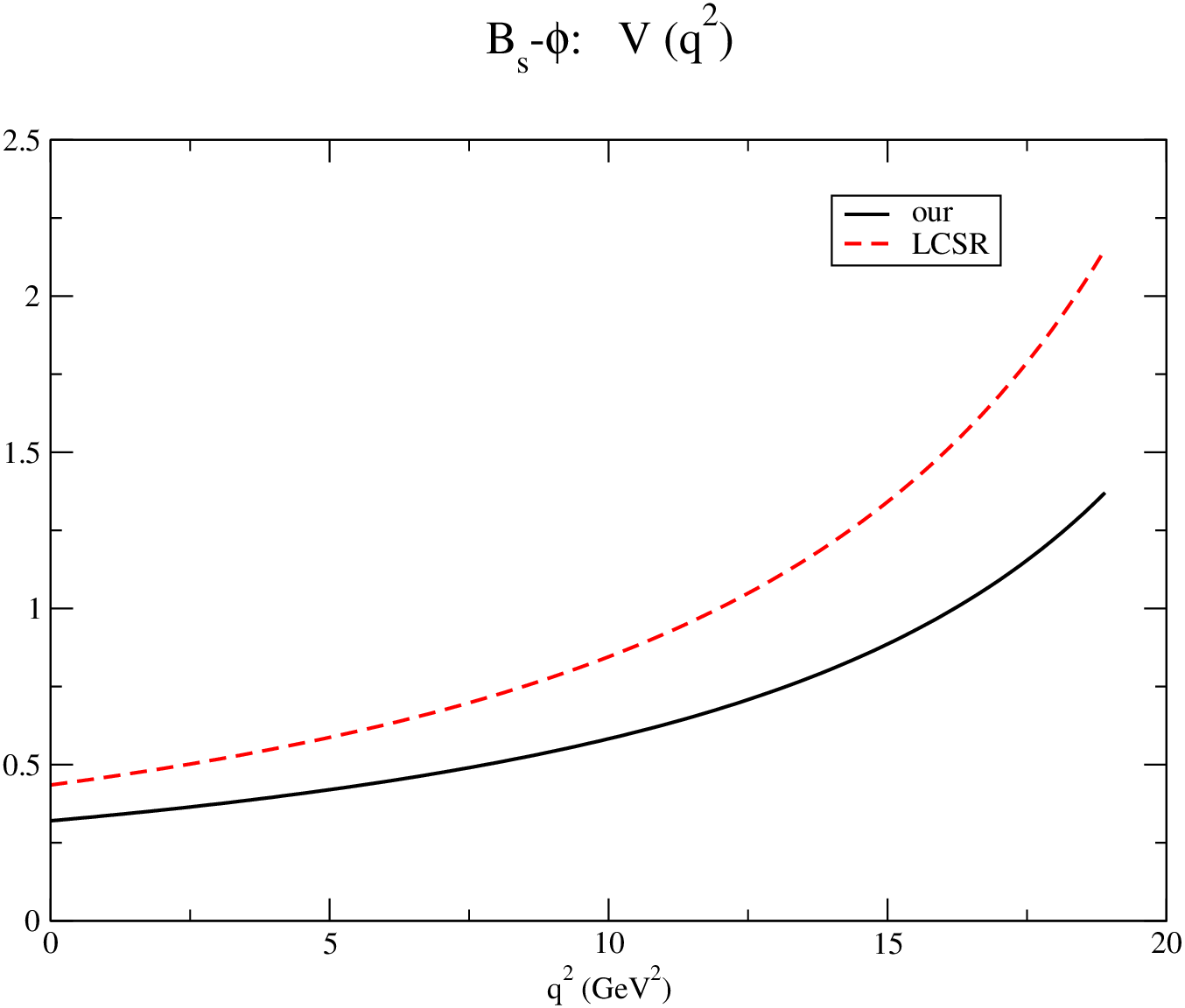} \\[2ex]
\includegraphics[width=0.40\textwidth]{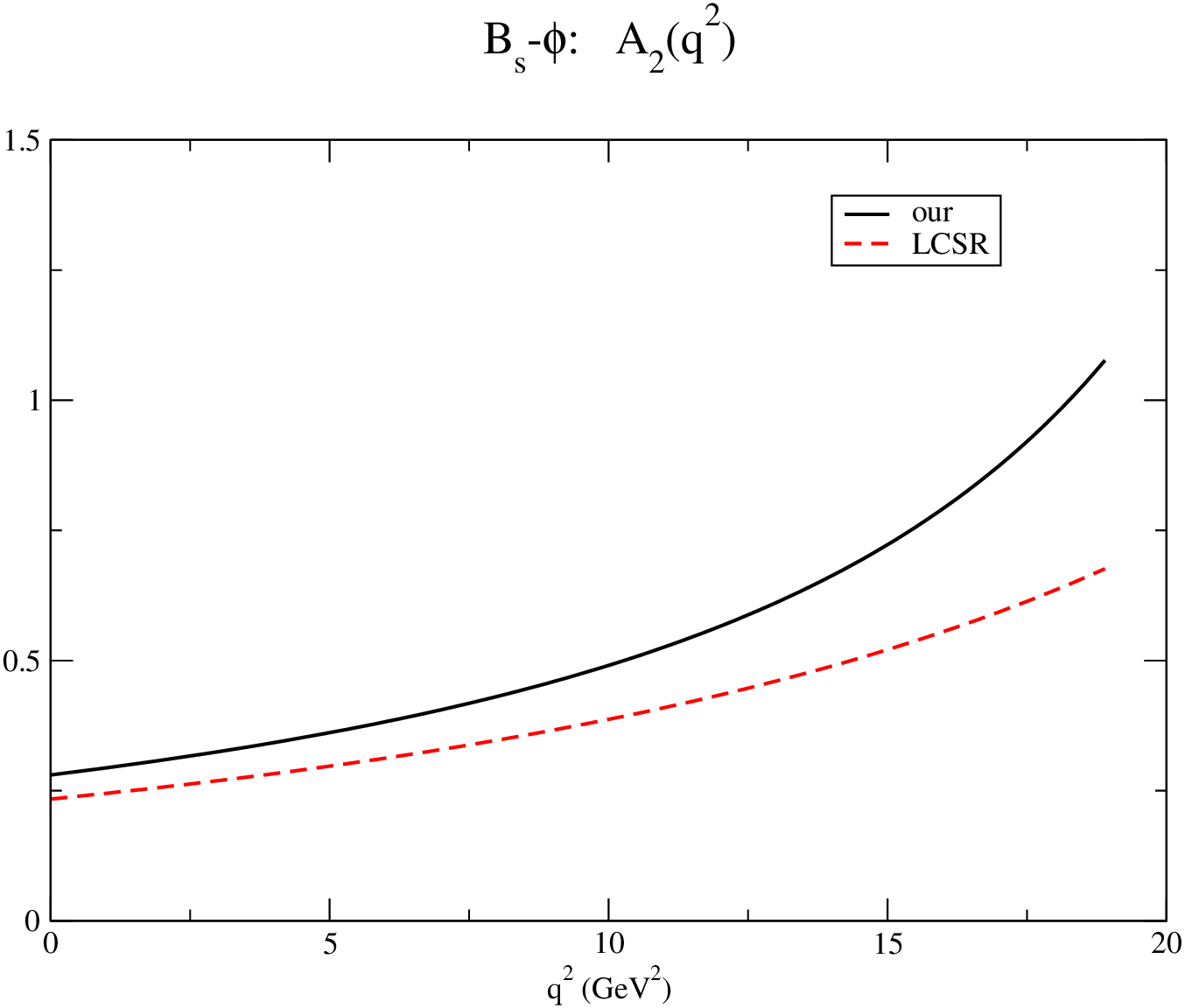} &
\includegraphics[width=0.40\textwidth]{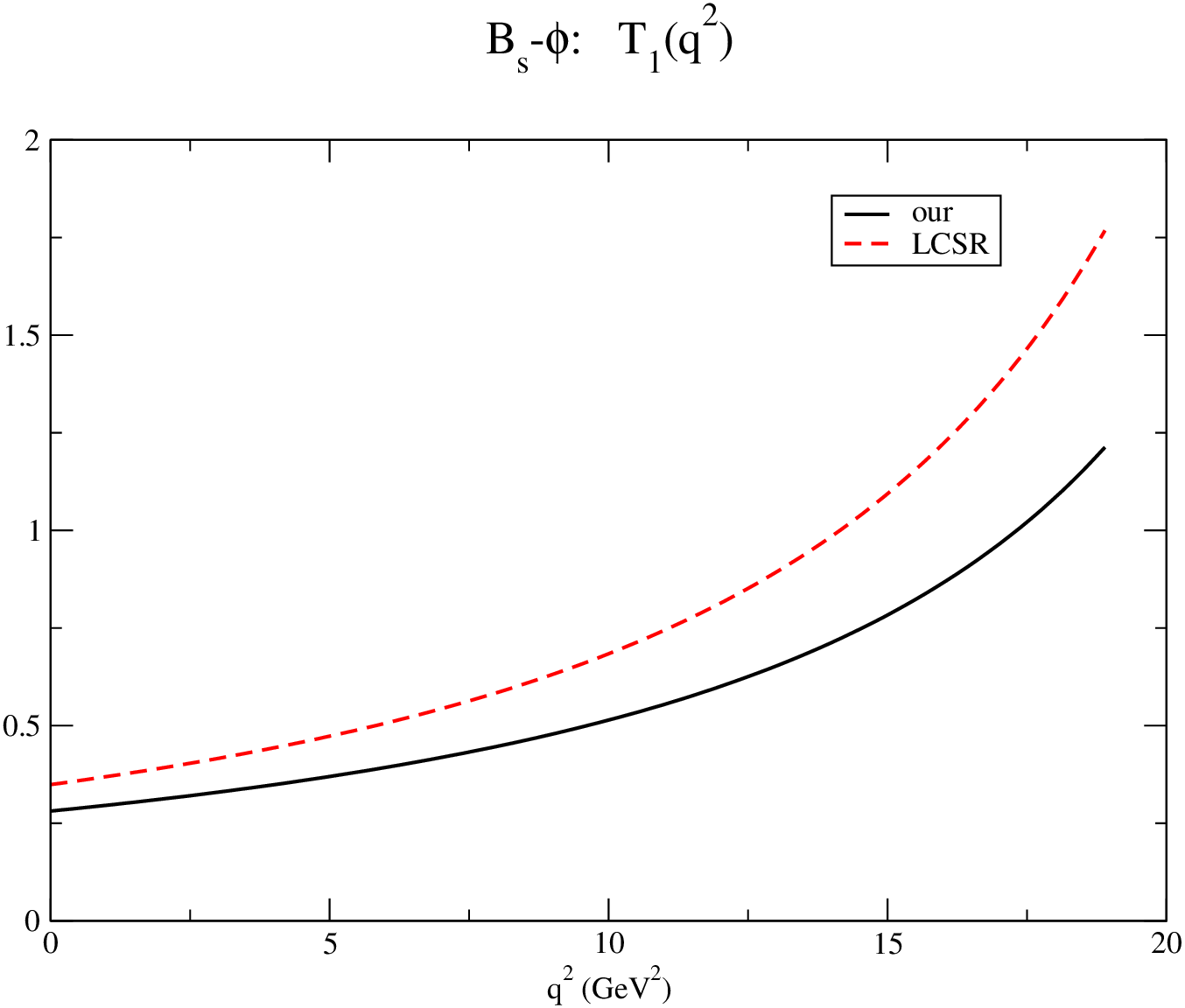}
\end{tabular}
\end{center}
\caption{\label{fig:ff-BsPhi}
Our results for the form factors appearing 
in Eqs.\,(\protect\ref{eq:PV}) 
      \& (\protect\ref{eq:PVT}) for  $B_s-\phi-$transition. 
For comporison we plot the curves given by LCSR
from Ref.~\protect\cite{Ball}. }
\end{figure}

In Figs.~\ref{fig:ff-BP}-\ref{fig:ff-BsPhi}
we plot our calculated form factors in the entire kinematical region
$0\le q^2\le q^2_{\rm max}$. For comporison we also show the results
obtained in the light-cone sum rules \cite{Ball}.
The figures highlight the wide range of phenomena accessible
within our approach.


\begin{figure}[ht]
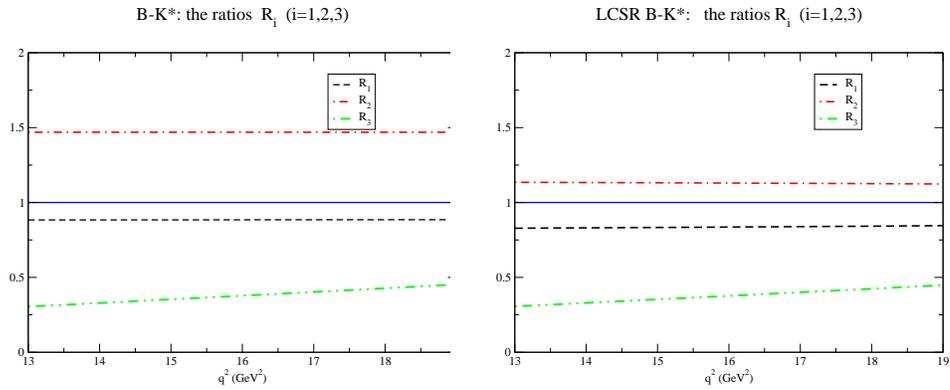

\begin{center}
\hspace*{-0.5cm}
\begin{tabular}{lr}
\includegraphics[width=0.40\textwidth]{BKv_ratio.eps} &
\includegraphics[width=0.40\textwidth]{BKv_LCSR_ratio.eps}
\end{tabular}
\end{center}
\caption{\label{fig:ff_ratio} Our results for the ratios 
of the form factors appearing 
in Eq.\,(\protect\ref{eq:ff_ratio}) 
 for  $B-K^\ast-$transition.} 
\end{figure}

As was suggested in Ref.~\cite{Hiller}, one can check
how well the form factors satisfy the low recoil relations
among them. In Fig.~\ref{fig:ff_ratio} we plot the ratios 
\begin{equation}
R_1 = \frac{T_1(q^2)}{V(q^2)}, \qquad
R_2 = \frac{T_2(q^2)}{A_1(q^2)}, \qquad
R_3 = \frac{q^2}{m^2_B}\frac{T_3(q^2)}{A_2(q^2)}.
\label{eq:ff_ratio}
\end{equation}
which in the symmetry limit should be all of order
$1-(2\alpha_s/(3\pi)\ln\left(\mu/m_b\right)$, i.e.
near one. One can see that similar to the LCSR form factors,
it works reasonably well for $R_1$ and $R_2$ but not for $R_3$.

\begin{figure}[ht]
\begin{center}
\includegraphics[width=0.60\textwidth]{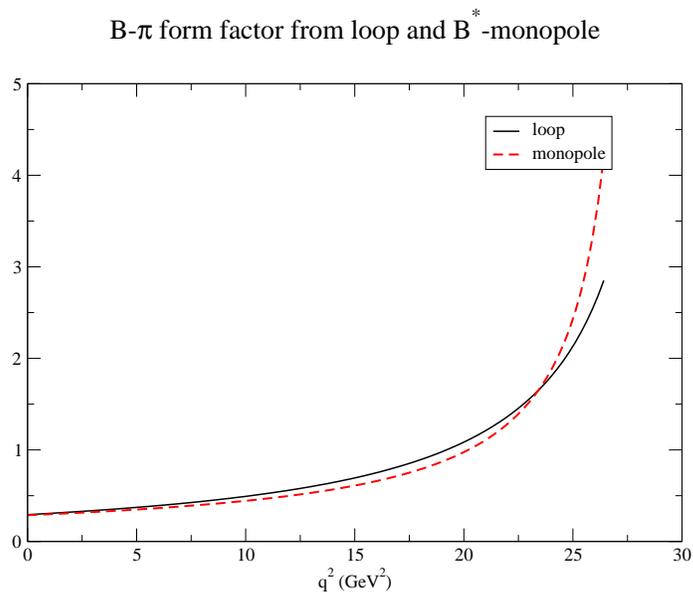} 
\end{center}
\caption{\label{fig:VDM} The comparison of the results for
the $B-\pi-$ form factor obtained on the one hand
from the quark-loop diagram and on the another hand
from the VDM-monopole.
}
\end{figure}

It is interesting to compare the behavior of the form factor
calculated from the triangle loop-diagram with those from
vector-dominance model (VDM). In the case of the $B-\pi-$transition,
one has
\[
F_{\rm VDM}^{B\pi}(q^2) = \frac{F_+^{B\pi}(0)}{m^2_{B^\ast}-q^2}.
\label{eq:VDM}
\] 
The curves are plotted in Fig.~\ref{fig:VDM}. One can see that
they agree with quite good accuracy. That means the quark loop
in some sense contains an information on the $B^\ast$-pole.

As an application of the obtained results we evaluate
the widths of the $B_s$-nonleptonic decays. 
The modes  
$D_s^- D_s^+,$  $D_s^{\ast\,-} D_s^{+}+D_s^- D_s^{\ast\,+}$ and
$D_s^{\ast\,-} D_s^{\ast\,+}$ give the largest contribution
to $\Delta\Gamma$ for the $B_s-\bar B_s$ system.
The mode $J/\psi\phi$ is suppressed by the color factor
but it is interesting for the search of CP-violating 
New-Physics possible effects in the $B_s-\bar B_s$ mixing.

For the CKM-matrix elements we use the values from \cite{PDG}
\begin{equation}
\def\arraystretch{2}
\begin{array}{ccccccc}
   |V_{ud}|    &    |V_{us}|     & |V_{ub}|       & |V_{cd}| & |V_{cs}| & |V_{cb}| \\
\hline
\ \ 0.974 \ \ & \ \ 0.225 \ \  & \ \ 0.00389\ \ & \ \ 0.230 \ \  & 
\ \ 0.975 \ \  & \ \ 0.0406 \ \ \\
\end{array}
\label{eq:CKM}
\end{equation}

For the Wilson coefficients we take
\cite{Altmannshofer:2008dz}
\begin{equation}
\def\arraystretch{2}
\begin{array}{ccccccc}
   C_1    &    C_2  & C_3   & C_4 & C_5 & C_6 \\
\hline
\ \ -0.257 \ \ & \ \ 1.009 \ \  & \ \ - 0.005\ \ & \ \ -0.078 \ \  & 
\ \ 0.000 \ \  & \ \ 0.001 \ \ \\
\end{array}
\label{eq:Wilson}
\end{equation}
evaluated to next-to-next-to leading logarithmic accuracy 
in $\overline{MS}$ (NDR) renormalization scheme at the scale
$\mu=4.8$~GeV \cite{Bobeth:1999mk}. 

We also need the values of the $B_s-\phi-$transition evaluated
at $q^2=m^2_{J/\psi}$. We give them in Table~\ref{tab:ff-BsPhi} 
and compare with results of Ref.~\cite{Faller:2008gt}.

\begin{table}[ht]
\begin{center}
\def\arraystretch{1.5}
     \caption{The relevant $B_s-\phi-$form factors at $q^2=m^2_{J/\psi}$
calculated in our work. For comparison we give the results of 
Ref.~\cite{Faller:2008gt}.  }
\label{tab:ff-BsPhi}
\begin{tabular}{lcc}
\hline\hline
  & This work & Ref.~\cite{Faller:2008gt} \\
\hline
$A_1(m^2_{J/\psi})$ & 0.37 & 0.42$\pm$0.06 \\
$A_1(m^2_{J/\psi})$ & 0.48 & 0.38$\pm$0.06 \\
$V(m^2_{J/\psi})$  & 0.56  & 0.82$\pm$0.12 \\
\hline\hline
\end{tabular}
\end{center}
\end{table}
Finally, we give our results for the branching ratios in
Table~\ref{tab:nonlep-widths}. One can see that there is good
agreement with available experimental data.

\begin{table}[ht]
\begin{center}
\def\arraystretch{1.5}
     \caption{Branching ratios ($\%$) of the $B_s$-nonleptonic decays
 calculated in our approach.}
\label{tab:nonlep-widths}
\vspace*{0.2cm}
\begin{tabular}{lcc}
\hline\hline 
Process &\,\,\, This work\,\,\, & Data~\cite{PDG}  \\
\hline
$B_s\to D_s^- D_s^+ $     & 1.65 & $1.04^{+0.29}_{-0.26}$ \\
$B_s\to D_s^- D_s^{\ast\,+}+ D_s^{\ast\,-} D_s^{+}$  & 2.40 & $2.8 \pm 1.0 $ \\
$B_s\to D_s^{\ast\,-} D_s^{\ast\,+} $ & 3.18 & $3.1 \pm 1.4$\\
$B_s\to J/\psi\phi$ & 0.14  & \,\,$0.14 \pm 0.05$\,\, \\
\hline\hline
\end{tabular}
\end{center}
\end{table}

\clearpage
\section{Light baryons}

Let us begin our discussion with the proton.
The coupling of a proton to its constituent quarks is described by 
the Lagrangian 
\begin{equation}
\label{eq:Lagr_str}
{\cal L}^{\,\rm p}_{\rm int}(x) = g_N \,\bar p(x)\cdot J_p(x) 
                           + g_N \,\bar J_p(x)\cdot p(x)\,,  
\end{equation}
where we make use of the same interpolating three-quark current 
$J_p(\bar J_p)$ as in Ref.~\cite{Ivanov:1996pz}
\begin{eqnarray}
J_p(x) &=& \int\!\! dx_1 \!\! \int\!\! dx_2 \!\! \int\!\! dx_3 \, 
F_N(x;x_1,x_2,x_3) \, J^{(p)}_{3q}(x_1,x_2,x_3)\,, \nonumber\\
J^{(p)}_{3q}(x_1,x_2,x_3) &=& 
\Gamma^A\gamma^5 \, d^{a_1}(x_1) 
\cdot[\epsilon^{a_1a_2a_3} \, u^{a_2}(x_2) \,C \, \Gamma_A \, u^{a_3}(x_3)]\,,
\nonumber\\
\label{eq:current}\\ 
\bar J_p(x) &=& \int\!\! dx_1 \!\! \int\!\! dx_2 \!\! \int\!\! dx_3 \, 
F_N(x;x_1,x_2,x_3) \, \bar J^{(p)}_{3q}(x_1,x_2,x_3)\,,
\nonumber\\
\bar J^{(p)}_{3q}(x_1,x_2,x_3) &=& 
[\epsilon^{a_1a_2a_3} \, 
\bar u^{a_3}(x_3)\, \Gamma_A \,C\, \bar u^{a_2}(x_2)]  
\cdot \bar d^{a_1}(x_1) \gamma^5\Gamma^A\,.
\nonumber
\end{eqnarray}
The matrix $C=\gamma^{0}\gamma^{2}$ is
the usual charge conjugation matrix and the $a_i$ $(i=1,2,3)$ are color 
indices. There are two possible kinds of nonderivative three-quark currents:
$\Gamma^A\otimes \Gamma_A = \gamma^\alpha\otimes \gamma_\alpha$~(vector
current) and 
$\Gamma^A\otimes \Gamma_A = \frac{1}{2} \, \sigma^{\alpha\beta}\otimes 
\sigma_{\alpha\beta}$~(tensor current) with  
$\sigma^{\alpha\beta}
= \frac{i}{2}(\gamma^\alpha\gamma^\beta-\gamma^\beta\gamma^\alpha).$ 
The interpolating current of the neutron and the corresponding Lagrangian are 
obtained from the proton case via $p \to n$ and 
$u \leftrightarrow d$. As will become apparent later on, one has to consider
a general linear superposition of the vector and tensor currents according to 
\begin{equation}
\label{eq:superpo}
J_N = x J^T_N + (1-x) J^V_N\,, \quad N = p, n 
\end{equation}

\begin{figure}[ht]
\begin{center}
\includegraphics[width=0.60\textwidth]{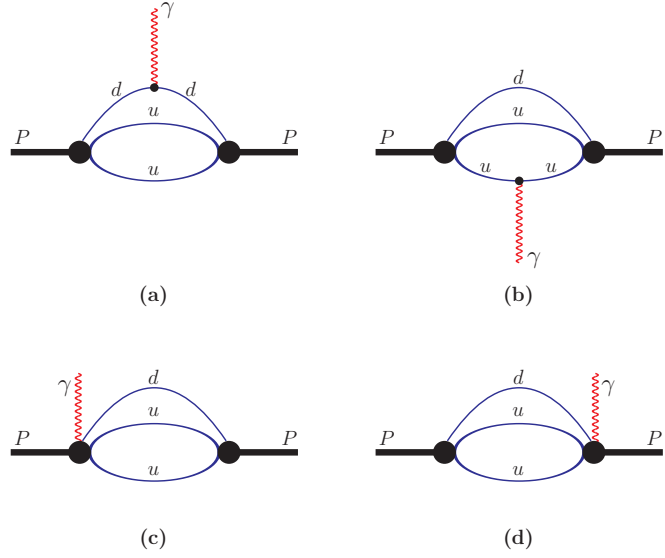} 
\end{center}
\caption{
\label{fig:em-proton}
Electromagnetic vertex function of the proton: 
(a) vertex diagram with the e.m. current attached to d-quark; 
(b) vertex diagram with the e.m. current attached to u-quark; 
(c) bubble diagram with the e.m. current attached to the initial state vertex; 
(d) the bubble diagram with e.m. current attached to the final state vertex.}   
\end{figure}

The electromagnetic vertex function $\Lambda_p^\mu(p,p')$ of the proton
consists of four pieces represented by the four two-loop quark diagrams 
in Fig.~\ref{fig:em-proton}.

Let us briefly describe a check on the gauge invariance of our 
calculation. Without gauge invariance there are three independent Lorentz 
structures in the electromagnetic proton vertex which can be chosen to be
\begin{equation}
\Lambda_p^\mu(p,p')=\gamma^\mu\,F^p_1(q^2) 
- \frac{i\sigma^{\mu q}}{2m_N}\,F^p_2(q^2)
+q^\mu\,F^p_{NG}(q^2)\,,
\label{eq:em-ff}
\end{equation}
where $\sigma^{\mu q} = 
\frac{i}{2}(\gamma^\mu\gamma^\nu-\gamma^\nu\gamma^\mu)q_\nu.$
The form factor $F^p_{NG}(q^2)$ characterizes the non--gauge invariant piece
and must therefore vanish for any $q^2$ in a calculation which respects
gauge invariance. For the four contributions of Fig.~2a-2d we found that 
\begin{equation}
F^p_{NG\, \rm d}(q^2) \equiv 0\,, \qquad  
F^p_{NG\, \rm u}(q^2) \equiv 0\,, \qquad 
F^p_{NG\, (b)}(q^2)   \equiv -\,F^p_{NG\, (a)}(q^2) \qquad \forall q^2.
\label{eq:NG-vanish}
\end{equation}  
It means that the non--gauge invariant contributions of 
the two vertex diagrams are zero while
they vanish for the sum of the two bubble diagrams. 

The electromagnetic vertex function of the neutron
is obtained from that of the proton by replacing $m_u\leftrightarrow m_d$,
$e_u\leftrightarrow e_d$ and $m_p\to m_n$. $F_1^N(q^2)$ and 
$F_2^N(q^2)$ are the Dirac and Pauli nucleon form factors 
which are normalized to the electric charge $e_N$ and anomalous 
magnetic moment $k_N$ ($k_N$ is given in units of the nuclear magneton 
$e/2m_p$), respectively, i.e. one has $F_1^N(0)=e_N$ and $F_2^N(0)=k_N$.  
In particular, one can analytically check by using the integration-by-part 
identity  that the Dirac form factor of the neutron is equal to zero at $q^2=0$.
  
The nucleon magnetic moments $\mu_N = F_1^N(0)+F_2^N(0)$ 
are known experimentally with high accuracy~\cite{PDG}
\begin{equation}
\mu^{\rm expt}_p = 2.79 \qquad 
\mu^{\rm expt}_n=-1.91 \,.
\label{eq:mag-mom-expt}
\end{equation}
We will use these values to fit the value of the nucleon size parameter.
We obtain 
\begin{eqnarray}
\mbox{\rm vector current} &\Longrightarrow&
\Lambda_N = 0.36\,  \mbox{\rm GeV}\, \quad \mu_p = 2.79 \quad  \mu_n = -1.70\,,
\\[2ex]
\mbox{\rm tensor current} &\Longrightarrow&
\Lambda_N = 0.61\,  \mbox{\rm GeV}\, \quad \mu_p = 2.79 \quad  \mu_n = -1.69\,.
\label{eq:mag-mom}
\end{eqnarray}

It is convenient to introduce the Sachs electromagnetic form factors 
of nucleons 
\begin{equation}
G_E^N(q^2) = F_1^N(q^2) + \frac{q^2}{4m^2_N} F_2^N(q^2)\,,
\qquad
G_M^N(q^2) = F_1^N(q^2) + F_2^N(q^2)\,.
\label{eq:GEGM}
\end{equation} 
The slopes of these form factors are related to the well-known 
electromagnetic radii of nucleons: 
\begin{equation}
\langle r^2_E \rangle^N = 6 \frac{dG_N^E(q^2)}{dq^2}\bigg|_{q^2 = 0} \,, 
\qquad
\langle r^2_M \rangle^N = \frac{6}{G_M^N(0)} \,
\frac{dG_M^N(q^2)}{dq^2}\bigg|_{q^2 = 0}  \,. 
\end{equation} 

We would like to emphasize that reproducing data on the neutron charge radius
$\langle r^2_E \rangle^n$ is a nontrivial task (see e.g. 
discussion in Ref.\cite{deAraujo:2003ke}). 
As well-known the naive nonrelativistic quark model based on SU(6)
spin-flavor symmetry implies $\langle r^2_E \rangle^n \equiv 0$. The dynamical
breaking of the SU(6) symmetry based on the inclusion of the quark spin-spin
interaction generates a nonvanishing value of $\langle r^2_E \rangle^n$.
From this point of view the
dominant contribution to the $\langle r^2_E \rangle^n $ comes from
the Pauli term:
\[
\langle r^2_E \rangle^n \simeq \frac{6}{4m^2_{N}} F_2^n(0) \,. 
\]

The experimental data on the nucleon
Sachs form factors in the space-like region $Q^2=-q^2\ge 0$
can be approximately described by the dipole approximation
\[
G^p_E(q^2)\approx \frac{G^p_M(q^2)}{1+\mu_p} 
\approx \frac{G^n_M(q^2)}{\mu_n}
\approx \frac{4m_N^2}{q^2}\frac{G^n_E(q^2)}{\mu_n} 
 \approx \frac{1}{\left(1-q^2/0.71\,{\rm GeV}^2\right)^2}\equiv D_N(q^2)\,.
\] 
According to present data the dipole approximation
works well up to 1 GeV$^2$ (with an accuracy of up to 25\%). For higher values
of $Q^2$ the deviation of the nucleon form factors from the dipole 
approximation becomes more pronounced.
In particular, the best description of magnetic moments, electromagnetic 
radii and form factors is achieved when we consider a superposition of the
$V$-- and $T$--currents of nucleons according to Eq.~(\ref{eq:superpo}) 
with $x = 0.8$. For the size parameter of the nucleon we take 
$\Lambda_N = 0.5$ GeV.

\begin{figure}[ht]
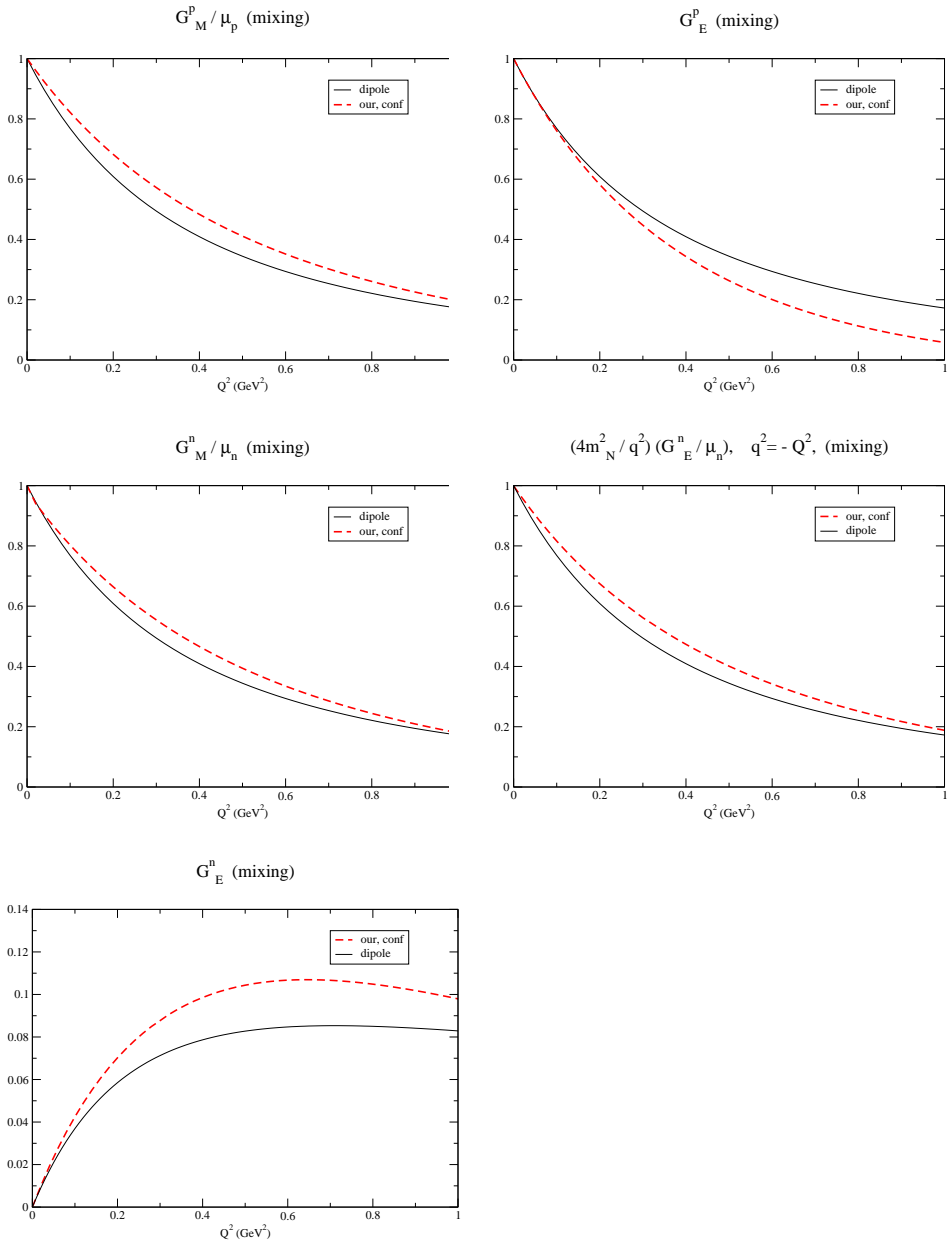

\begin{center}
\hspace*{-0.5cm}
\begin{tabular}{lr}
\includegraphics[width=0.40\textwidth]{GMp.conf.mix.eps} & 
\includegraphics[width=0.40\textwidth]{GEp_N.conf.mix.eps}
\\[2ex]
\includegraphics[width=0.40\textwidth]{GMn.conf.mix.eps} & 
\includegraphics[width=0.40\textwidth]{GEn_N.conf.mix.eps}
\\[2ex]
\includegraphics[width=0.40\textwidth]{GEn.mix.eps} &
\end{tabular}
\end{center}
\caption{
\label{fig:em-ff}
Sachs nucleon form factors in comparions with the dipole 
representation in the space--like region $Q \le 1$ GeV$^2$.}
\end{figure}

In Table~\ref{tab:nuc_res} we present the results for the magnetic moments and 
electromagnetic radii for this set of model parameters. 
In Fig.~\ref{fig:em-ff} we present our results for the $q^{2}$ 
dependence of electromagnetic  form factors in the 
region $Q^2\in [0,1]\,{\rm GeV}^2$. Fig.~\ref{fig:em-ff} also shows plots of 
the dipole  approximation to the form factors. 
The agreement of our results with the 
dipole approximation is satisfactory. Inclusion of chiral corrections as,
for example, developed and discussed in~\cite{Faessler:2005gd} may lead to a
further improvement in the low $Q^{2}$ description.  

\begin{table}[ht]
\begin{center}
\caption{
\label{tab:nuc_res}
Electromagnetic properties of nucleons.}
\def\arraystretch{1.5}
    \begin{tabular}{|c|c|c|}
      \hline
Quantity & Our results & Data~\cite{PDG}  \\
\hline
$\mu_p$ (in n.m.)          &  2.96       &  2.793              \\
\hline
$\mu_n$ (in n.m.)          & -1.83       & -1.913              \\
\hline
$r_E^p$ (fm)     &  0.805 &  0.8768 $\pm$ 0.0069 \\
\hline
$\langle r^2_E \rangle^n$ (fm$^2$) & -0.121 & -0.1161 $\pm$ 0.0022 \\
\hline
$r_M^p$ (fm)     &  0.688 &  0.777  $\pm$ 0.013 $\pm$ 0.010 \\
\hline
$r_M^n$ (fm)     &  0.685 &  0.862$^{+0.009}_{-0.008}$     \\
\hline
\end{tabular}
\end{center}
\end{table}

\clearpage
\section{The X(3872)-meson  as a tetraquark}

A narrow charmonium--like state $X(3872)$ was observed in 2003 
in the exclusive decay process $B^\pm\to K^\pm\pi^+\pi^-J/\psi$
\cite{Choi:2003ue}. 
The  $X(3872)$ decays into $\pi^+\pi^-J/\psi$ and has a mass of 
$m_X=3872.0 \pm 0.6 ({\rm stat}) \pm 0.5 ({\rm syst}) $
very close to the $M_{D^0}+M_{D^{\ast\,0}}=3871.81 \pm 0.25$ mass threshold 
\cite{PDG}.
Its width was found to be less than 2.3 MeV at $90\%$ confidence level.
The state was confirmed in B-decays by the BaBar experiment 
\cite{Aubert:2004fc} 
and in $p\overline{p}$ production
by the Tevatron experiments \cite{Tevatron}. 

From the  observation of the decay $X(3872)\rightarrow J/\psi \gamma$ reported 
by \cite{jpsigamma}, 
it was shown that the only quantum numbers 
compatible with the data are $J^{PC}=1^{++}$ or $2^{-+}$. 
However, the observation of the decays into 
$D^0\overline{D}^{0}\pi^0$ by the Belle and BaBar collaborations 
\cite{jpsiDD} allows one to exclude the choice $2^{-+}$
because the near-threshold decay 
$X\to D^0\overline{D}^{0}\pi^0$   is expected to be strongly suppressed 
for $J=2$.

The Belle collaboration has  reported evidence for the decay mode  
$X \to \pi^+\pi^-\pi^0 J/\psi$ with  a strong three-pion peak between 750 MeV
and the kinematic limit of 775 MeV \cite{jpsigamma}, suggesting that 
the process is dominated by the  sub-threshold decay $X \to \omega J/\psi$. 
It was found that the branching ratio
of this mode is almost the same as that of the mode $X \to \pi^+\pi^- J/\psi$:
\begin{equation}
\hspace*{-0.5cm}
\frac{ {\cal B}(X\to J/\psi\pi^+\pi^-\pi^0) }
     { {\cal B}(X\to J/\psi\pi^+\pi^-) }
 = 1.0 \pm 0.4 ({\rm stat}) \pm 0.3 ( {\rm syst} ).
\label{eq:ratio-expt}
\end{equation}
These observations imply strong isospin violation because the three-pion decay
proceeds via an intermediate $\omega$-meson with isospin 0 whereas
the two-pion decay proceeds via the intermediate $\rho$-meson with isospin 1.
Also the two-pion decay via the intermediate $\rho$-meson is very difficult
to explain by using an interpretation of the $X(3872)$ as a
simple $c\bar c$ charmonium state with isospin 0.  

There are several different interpretations of the $X(3872)$ in the literature:
a molecule bound state ($D^0\overline{D}^{\ast\,0}$)  with small binding energy, 
a tetraquark state composed of a diquark and antidiquark,
threshold cusps,  hybrids  and glueballs.
A description of the current theoretical and experimental situation for the 
new charmonium states may be found in the reviews \cite{review-X}. 

We provided in Ref.~\cite{Dubnicka:2010kz} an independent analysis of 
the properties of the $X(3872)$ meson which we interpret as 
a tetraquark state as in \cite{Maiani:2004vq}.
The authors of \cite{Maiani:2004vq} suggested to consider the $X(3872)$ meson
as a $J^{PC}=1^{++}$ tetraquark state with a symmetric spin distribution:
$[cq]_{S=0}\,[\bar c \bar q]_{S=1} + [cq]_{S=1}\,[\bar c \bar q]_{S=0}$,
$(q=u,d)$. The nonlocal version of the four-quark interpolating current
reads
\begin{eqnarray}
J^\mu_{X_q}(x) &=& \int\! dx_1\ldots \int\! dx_4 
\delta\left(x-\sum\limits_{i=1}^4 w_i x_i\right) 
\,\Phi_X\Big(\sum\limits_{i<j} (x_i-x_j)^2 \Big)
\nonumber\\
&\times&
\frac{1}{\sqrt{2}}\, \varepsilon_{abc}\varepsilon_{dec} \,
\Big\{\, [q_a(x_4)C\gamma^5 c_b(x_1)][\bar q_d(x_3)\gamma^\mu C \bar c_e(x_2)]
 + (\gamma^5 \leftrightarrow \gamma^\mu)
\,\Big\},
\label{eq:cur}
\end{eqnarray} 
where $w_1 = w_2 = m_c/2(m_q+m_c)$
and
$w_3 = w_4 = m_q/2(m_q+m_c)$.
The matrix $C=\gamma^0\gamma^2$ is the charge conjugation matrix.
The effective interaction Lagrangian describing the coupling
of the meson $X_q$ to its constituent quarks is written in the form
\begin{equation}
{\cal L}_{\rm int} = g_X\,X_{q\,\mu}(x)\cdot J^\mu_{X_q}(x), \qquad (q=u,d).
\label{eq:lag}
\end{equation}     
The state $X_u$ breaks isospin symmetry maximally so
the authors of \cite{Maiani:2004vq} take the physical states to be a linear
superposition of the $X_u$ and $X_d$ states according to
\begin{eqnarray}  
X_l\equiv X_{\rm low} &=&\hspace{0.2cm}  X_u\, \cos\theta +  X_d\, \sin\theta,\nonumber\\
X_h\equiv X_{\rm high} &=& - X_u\, \sin\theta +  X_d\, \cos\theta.
\label{eq:mixing}
\end{eqnarray}
The mixing angle $\theta$ can be determined from fitting the ratio
of branching ratios Eq.~(\ref{eq:ratio-expt}).

\begin{figure}[htbp]
\begin{center}
\includegraphics[scale=0.5]{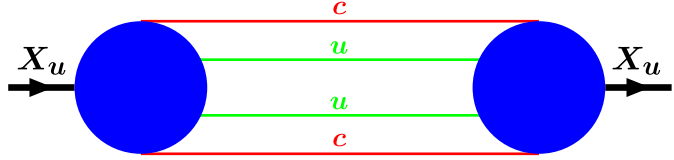} 
\caption{
\label{fig:X-mass}
Diagram describing the $X_{u}$-meson mass operator.}
\end{center}
\end{figure}

The coupling constant $g_X$ in Eq.~(\ref{eq:lag}) will be determined from
the compositeness condition:

\[
Z_X = 1-\Pi_X^\prime(m^2_X)=0,
\]
where $\Pi_X(p^2)$ is the scalar part of the vector-meson mass operator.
The corresponding three-loop diagram describing the X-meson mass operator is 
shown in  Fig.~\ref{fig:X-mass}.

\begin{figure}[htbp]
\begin{center}
\begin{tabular}{lr}
\includegraphics[scale=0.4]{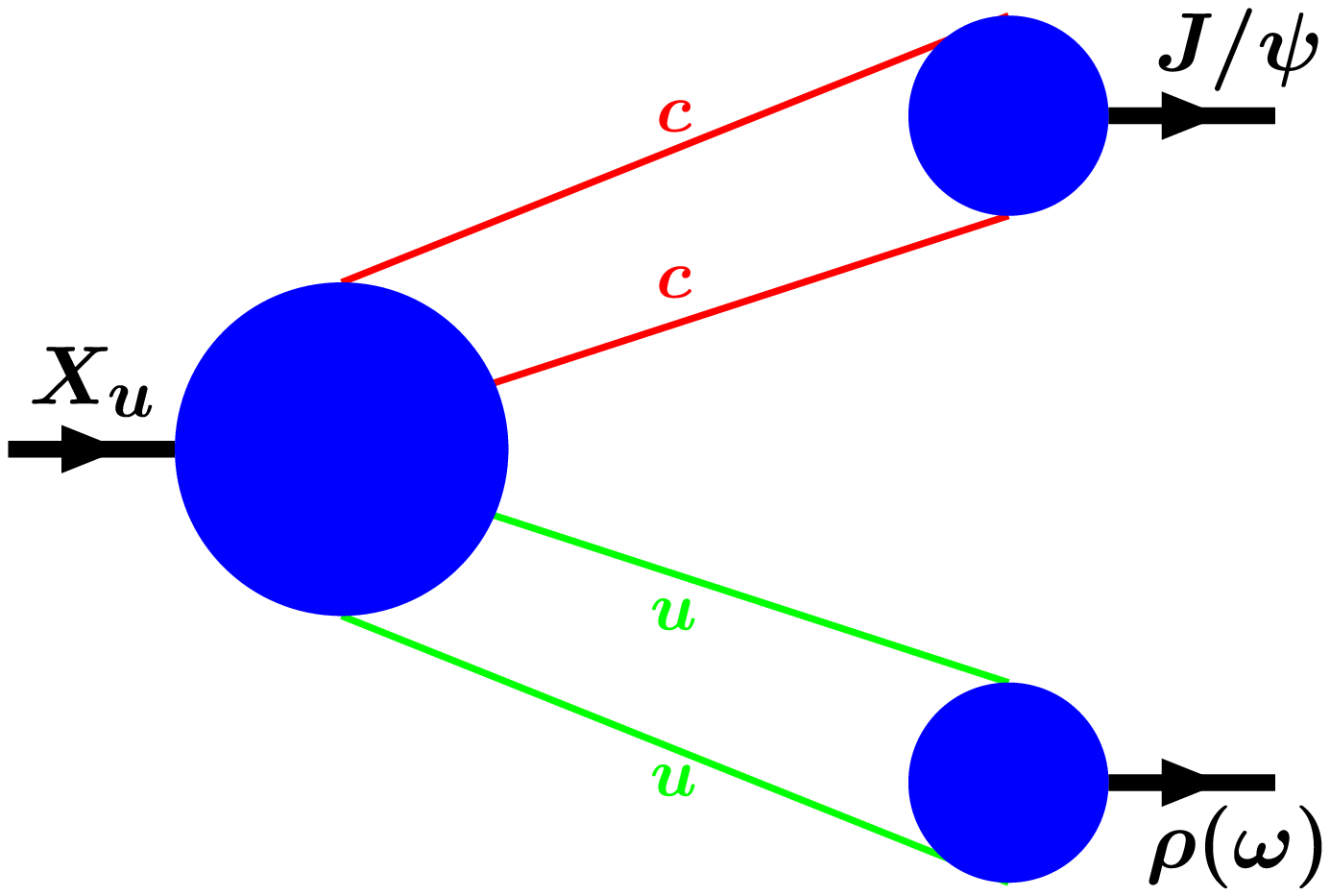} &
\includegraphics[scale=0.4]{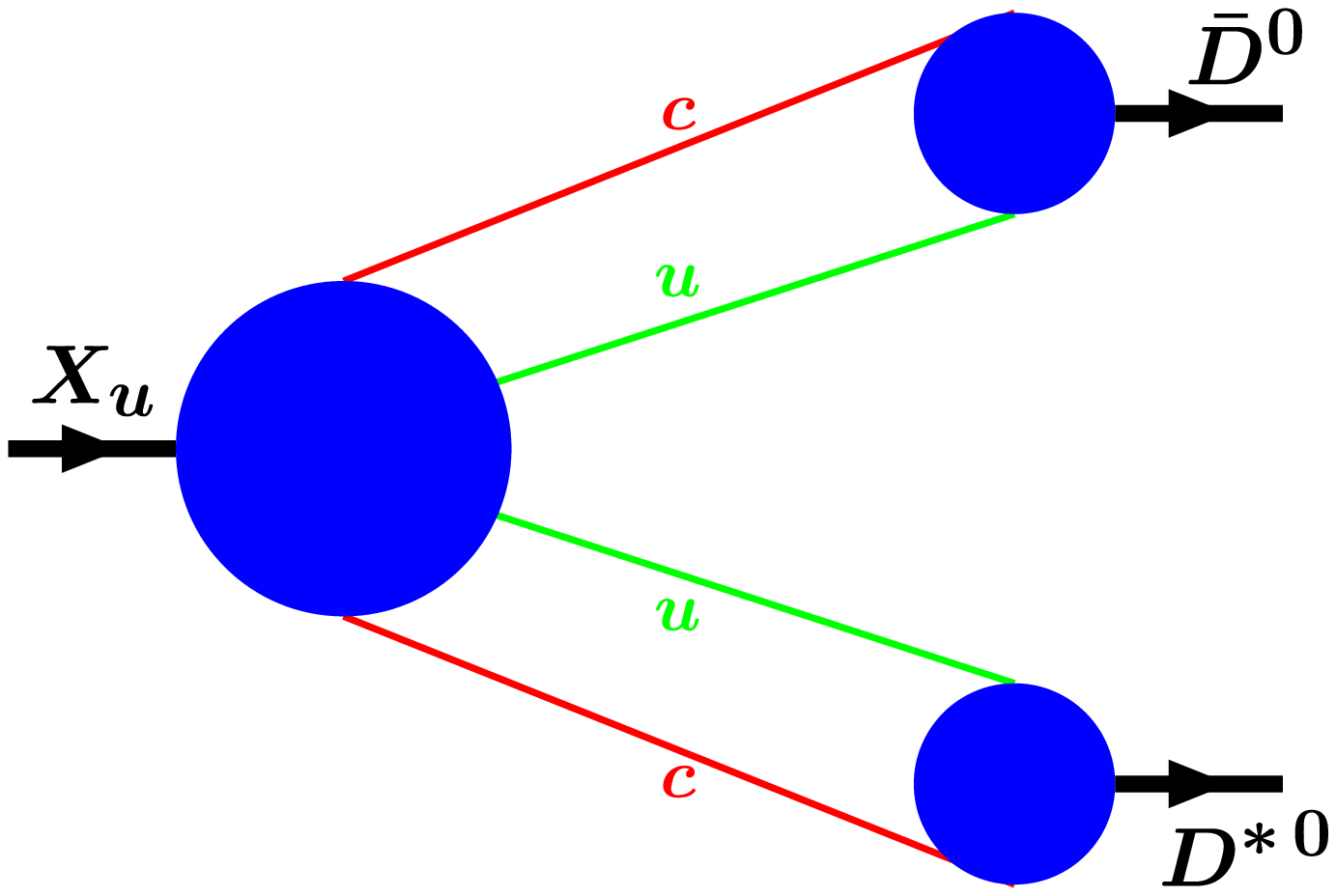}
\end{tabular}
\end{center}
\caption{
\label{fig:X-decay}
Feynman diagrams describing the decays
$X\to J/\psi+\rho(\omega)$ and $X\to D+\bar D^\ast$.
}
\end{figure}

Next we evaluate the matrix elements of the transitions
$X\to J/\psi+\rho(\omega)$ and $X\to D+\bar D^\ast$. 
The relevant Feynman diagrams are shown in Fig.~\ref{fig:X-decay}.

Since the X(3872) is very close to the respective thresholds in both cases, 
the intermediate $\rho$, $\omega$ and $D^\ast$ mesons 
have to be treated as off-shell particles.
Using the calculated matrix elements for the decay 
$X\to J/\psi+\rho(\omega)$ one can evaluate the decay widths
$X\to J/\psi+2\pi(3\pi) $. We employ the narrow width approximation
for this purpose.

There are two new free parameters: the mixing angle
$\theta$ in Eq.~(\ref{eq:mixing}) and the size parameter 
$\Lambda_X$. We have varied the parameter $\Lambda_X$
in a large interval and found that the ratio
\[
\frac{\Gamma(X_u\to J/\psi+3\,\pi)} 
     {\Gamma(X_u\to J/\psi+2\,\pi)} \approx 0.25
\]  
is very stable under variations of $\Lambda_X$. Hence, by using this result and
the central value of the experimental data given in Eq.~(\ref{eq:expt}),
one finds $\theta\approx \pm 18.4^{\rm o}$ for $X_l$~("+") and  $X_h$~("-"),
respectively. This is in agreement with the results obtained in both
\cite{Maiani:2004vq}: $\theta\approx \pm 20^{\rm o}$ and
\cite{Navarra:2006nd}: $\theta\approx \pm 23.5^{\rm o}$.
The decay width is quite sensitive to the change of the size
parameter $\Lambda_X$. A natural choice is to take a value 
close to $\Lambda_{J/\psi}$ and  $\Lambda_{\eta_c}$ 
which are both around 3 GeV.
We have varied the size parameter $\Lambda_X$ from  2.4 up to 4 GeV 
and found that
the decay width $\Gamma(X\to J/\psi+n\,\pi)$ decreases  from  0.25~MeV
monotonously. This result is in accordance with 
the experimental bound  $\Gamma(X(3872))\le 2.3$~MeV and the result
obtained in \cite{Maiani:2004vq}: 1.6~MeV.

\begin{figure}[ht]
\vspace*{0.5cm}
\begin{center}
\includegraphics[scale=0.35]{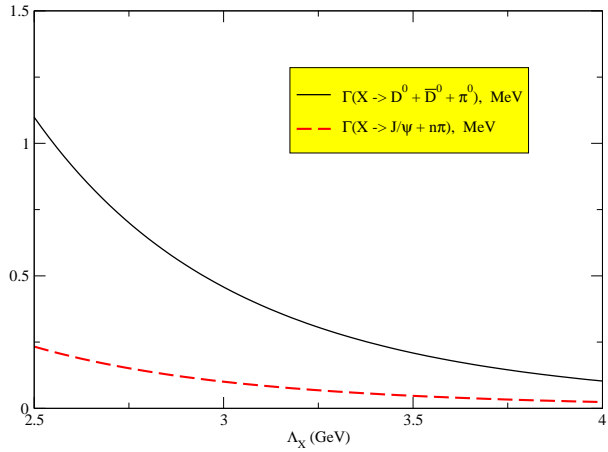}
\end{center}
\caption{
\label{fig:X-width}
The dependence of the decay widths 
$\Gamma(X_l\to \bar D^0 D^0 \pi^0)$
and $\Gamma(X\to J/\psi+n\pi)$ on the size parameter $\Lambda_X$.}
\end{figure}

In a similar way we calculate the width of the decay 
$X\to D^0\bar D^0\pi^0 $ which was observed by the Belle Coll.
and reported in \cite{jpsiDD}. 
As in the previous case we have  varied  $\Lambda_X$ from  2.5 up to 4 GeV 
and found that the decay width $\Gamma(X_l\to \bar D^0 D^0 \pi^0)$ decreases  
from  1.1~MeV monotonously. We plot  the dependence of 
the calculated decay widths on the size parameter $\Lambda_X$
in Fig.~\ref{fig:X-width}.

\vspace*{0.5cm}

Using the results of \cite{PDG}, one calculates the experimental rate ratio

\begin{equation}
\frac{\Gamma(X\to D^0\bar D^0 \pi^0)}
     {\Gamma(X\to J/\psi\pi^+\pi^-)}  = 10.5\pm 4.7
\label{eq:ratio}
\end{equation}
The theoretical value for this rate ratio depends only weakly on the size 
parameter $\Lambda_{X}$:

\begin{equation}
\frac{\Gamma(X\to D^0\bar D^0 \pi^0)}
     {\Gamma(X\to J/\psi\pi^+\pi^-)}\Big|_{\rm theor}  = 4.5 \pm 0.2.
\end{equation}
The theoretical error reflects the $\Lambda_{X}$ dependence of the 
ratio. The ratio lies within the experimental
uncertainties given by Eq.~(\ref{eq:ratio}).

\begin{figure}[ht]
\begin{center}
\begin{tabular}{lr}
\includegraphics[scale=0.4]{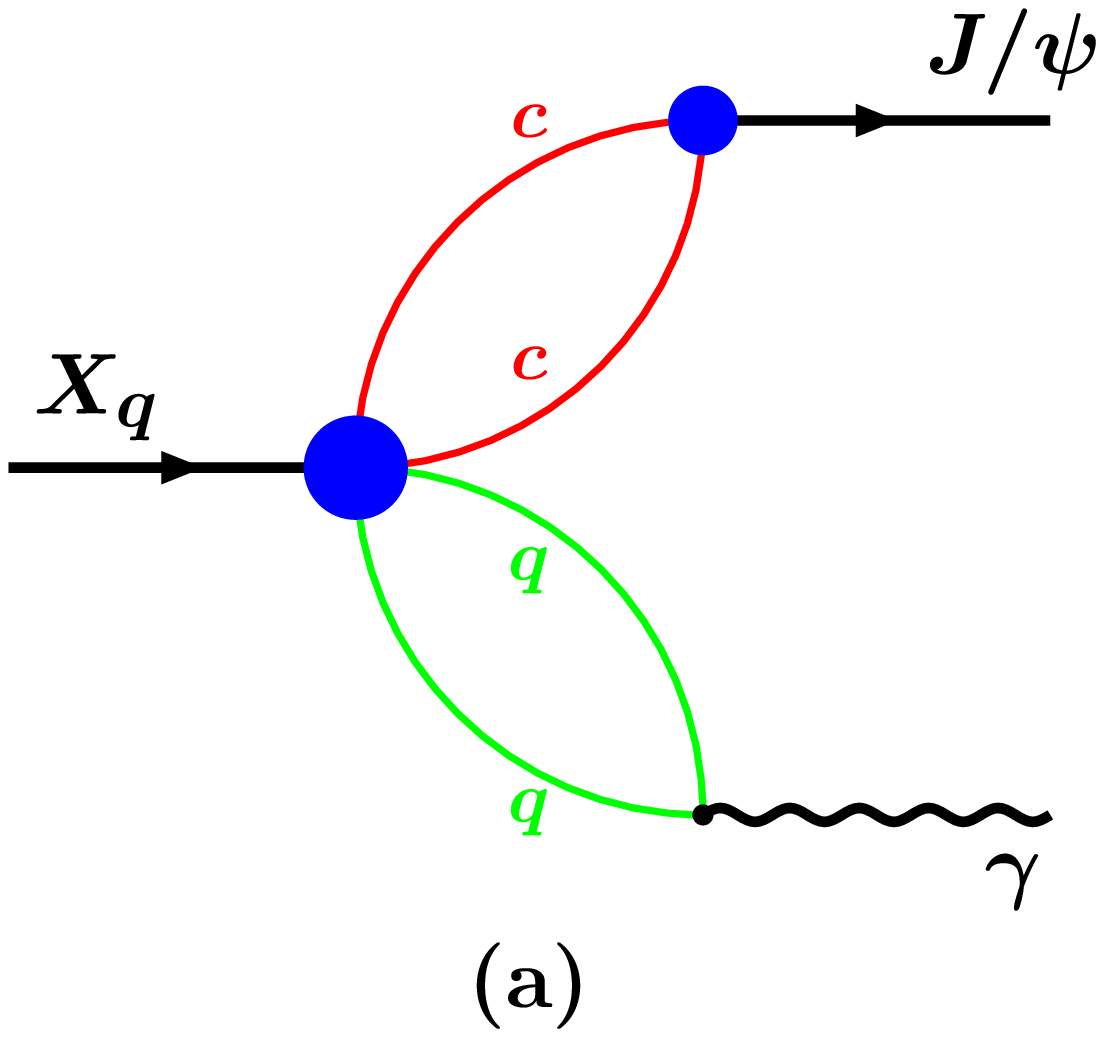} & 
\includegraphics[scale=0.4]{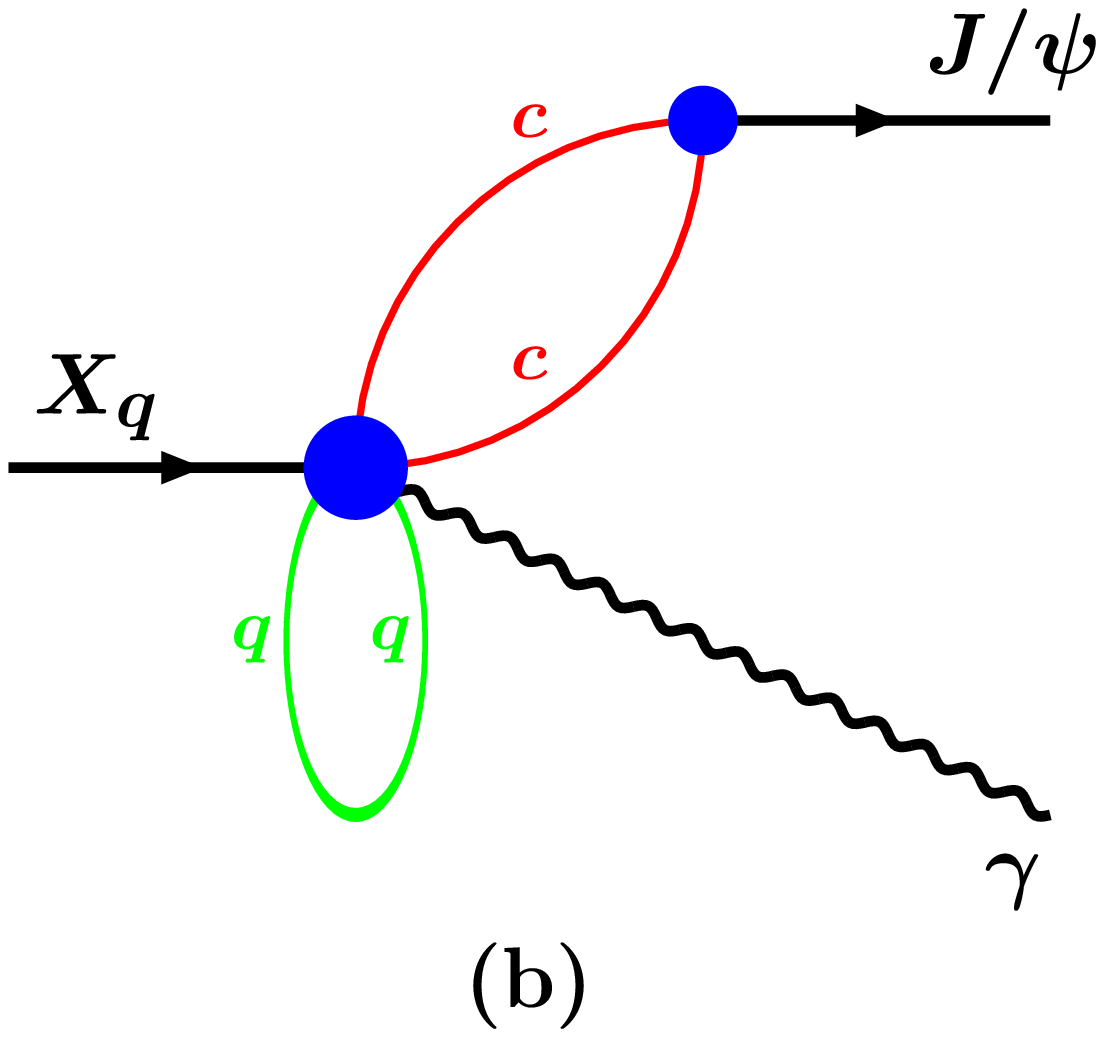}\\[2ex]
\includegraphics[scale=0.4]{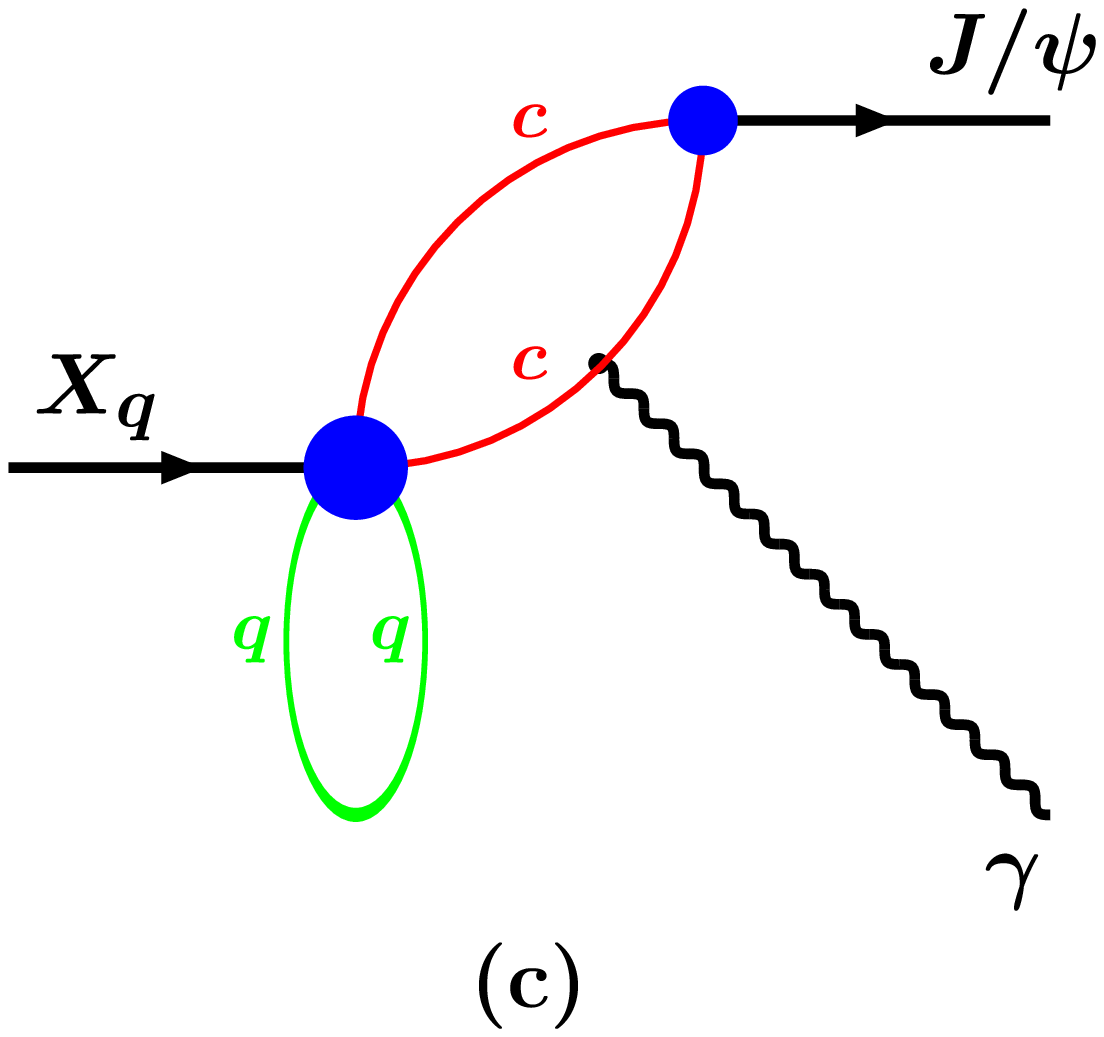} & 
\includegraphics[scale=0.4]{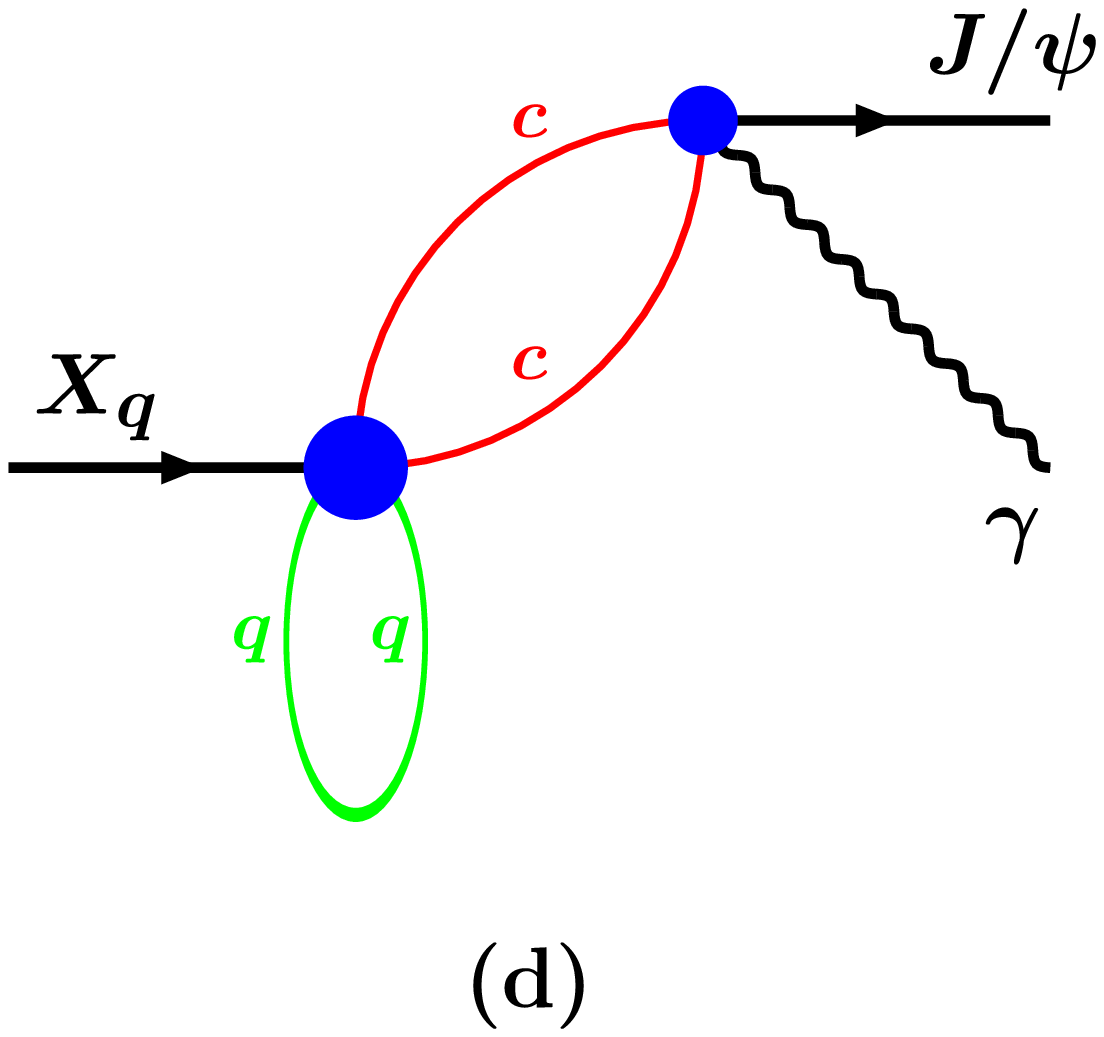}
\end{tabular}
\end{center}
\caption{
\label{fig:X-em-decay}
Feynman diagrams describing the decay
$X\to J/\psi+\gamma$.
}
\end{figure}

The matrix element of the decay
$X(3872)\to J/\psi+\gamma$ can be calculated from 
the Feynman diagrams shown in Fig.~\ref{fig:X-em-decay}.

The invariant matrix element for the decay is given by
\begin{equation}
M(X_q(p)\to J/\psi(q_1)+\gamma(q_2)) =
i(2\pi)^4\delta^{(4)}(p-q_1-q_2)\,
\varepsilon_X^\mu\,\varepsilon^\rho_\gamma\,\varepsilon^\nu_{J/\psi}\,
T_{\mu\rho\nu}(q_1,q_2)
\label{eq:matrix-element}
\end{equation}

We have analytically checked on the gauge invariance of the unintegrated 
transition matrix element by contraction with the photon momentum $q_{2}$
which yields $q^\rho_2  T_{\mu\rho\nu}(q_1,q_2)=0$ using the identities
\begin{eqnarray*}
&&
S(k_2)\!\not\! q_2 \,S(k_2+q_2) =  S(k_2+q_2) - S(k_2)\,,
\\
&&
\int\limits_0^1 d\tau\,
\widetilde\Phi^\prime(-\tau\,a-(1-\tau)\,b)\,(a-b)=
\widetilde\Phi(-b)-\widetilde\Phi(-a).
\end{eqnarray*}
If one takes the on-mass shell conditions
\begin{equation}
\varepsilon_X^\mu  p_\mu=0, \qquad \varepsilon_{J/\psi}^\nu q_{1 \nu}=0, \qquad
\varepsilon_\gamma^\rho q_{2 \rho}=0
\label{eq:on-mass-shell}
\end{equation}
into account one can write down five seemingly independent Lorentz structures
\[
T_{\mu\rho\nu}(q_1,q_2) = \varepsilon_{q_2\mu\nu\rho} q_1^2\,W_1
                      +\varepsilon_{q_1q_2\nu\rho} q_{1 \mu}\,W_2
                      +\varepsilon_{q_1q_2\mu\rho} q_{2 \nu}\,W_3
                      +\varepsilon_{q_1q_2\mu\nu} q_{1 \rho}\,W_4
                      +\varepsilon_{q_1\mu\nu\rho} q_1q_2\,W_5\,.
\]
Further, using the gauge invariance condition 
\[
q_2^\rho T_{\mu\rho\nu}=q_1q_2\varepsilon_{q_1q_2\mu\nu}(W_{4}+W_{5})
=0
\]
one has $W_{4}=-W_{5}$ which reduces the set of independent covariants to four:
\[
T_{\mu\rho\nu}(q_1,q_2) = \varepsilon_{q_2\mu\nu\rho} q_1^2\,W_1
                      +\varepsilon_{q_1q_2\nu\rho} q_{1 \mu}\,W_2
                      +\varepsilon_{q_1q_2\mu\rho} q_{2 \nu}\,W_3
 +\Big(
   \varepsilon_{q_1q_2\mu\nu} q_{1 \rho} - q_1q_2\varepsilon_{q_1\mu\nu\rho}\,W_4
   \Big)\,.
\]
The gauge invariance condition $W_{4}=-W_{5}$
provides for a numerical check on the gauge invariance of our 
calculation as described further on. 

However, there are two nontrivial relations among the four covariants 
which can be derived by noting \cite{Korner:2003zq} that the tensor
\begin{equation}
T_{\mu[\nu_{1}\nu_{2}\nu_{3}\nu_{4}\nu_{5}]}=
g_{\mu\nu_{1}}\varepsilon_{\nu_{2}\nu_{3}\nu_{4}\nu_{5}}
+{\rm cycl.}(\nu_{1}\nu_{2}\nu_{3}\nu_{4}\nu_{5})
\end{equation}
vanishes in four dimensions since it is totally antisymmetric in the five 
indices $(\nu_{1},\nu_{2},\nu_{3},\nu_{4},\nu_{5})$. Upon contraction with
$q_{1}^{\mu}q_{1}^{\nu_{1}}q_{2}^{\nu_{2}}$ and 
$q_{2}^{\mu}q_{1}^{\nu_{1}}q_{2}^{\nu_{2}}$ one finds
\begin{eqnarray*}
&&
  q_1^2 \varepsilon_{q_2\mu\nu\rho}
+ \varepsilon_{q_1q_2\nu\rho} q_{1 \mu}
+\Big(\varepsilon_{q_1q_2\mu\nu} q_{1 \rho} - q_1q_2 \varepsilon_{q_1\mu\nu\rho}\Big)
=0\,
\\
&&
q_1q_2\varepsilon_{q_2\mu\nu\rho}
-\varepsilon_{q_1q_2\nu\rho} q_{1 \mu}
-\varepsilon_{q_1q_2\mu\rho} q_{2 \nu}
=0\,.
\end{eqnarray*}
It reduces the set of independent covariants to two.
This is the appropriate number of independent covariants since the photon
transition is described by two independent amplitudes as e.g. by the $E1$ and
$M2$ transition amplitudes.

The quantities $W_i$ are represented by the four-fold integrals
\begin{equation}
W_i=\int\limits_0^\infty\! dt\! \int\limits_0^1\! d^3 \beta\, 
F_{i}(t,\beta_1,\beta_2,\beta_3)
\label{eq:structures}
\end{equation}
where we have suppressed the additional dependence of the integrand $F_{i}$
on the set of variables $p^2,q_1^2,q_2^2;m_q,m_c,s_X,s_{J/\psi}$ with
$s_X=1/\Lambda_X^2$ and $s_{J/\psi}=1/\Lambda_{J/\psi}^2$. 
The integrals in Eq.~(\ref{eq:structures})  have branch points
at $p^2=4(m_q+m_c)^2$ (diagram in Fig.~\ref{fig:X-em-decay}-a)
and at  $p^2=4 m_c^2$ (diagrams in Figs.~\ref{fig:X-em-decay}-b,c,d).
At these points the integrals become divergent in the convential sense
when $t\to \infty$. Under numerical check on gauge invariance 
of the amplitude $T_{\mu\rho\nu}(q_1,q_2)$,
we assume that the X-meson momentum squared is below the nearest
unitarity threshold, i.e. $p^2<4m_c^2$. The gauge invariance condition is
independent of the overall couplings $g_X$ and $g_{J/\psi}$ and thus the
numerical check can be done irrelevant of their values.

In the next step we introduce an infrared cutoff $1/\lambda^2$ on the upper 
limit of the t-integration in Eq.~(\ref{eq:structures}). In this manner one 
removes all possible singularities and thereby guarantees quark confinement.
However, the contributions coming from the
bubble diagrams in Figs.~\ref{fig:X-em-decay}-b,c,d blow up at $p^2=m^2_X$ 
compare with the contribution from the diagram in Fig.~\ref{fig:X-em-decay}-a. 
The bubble diagrams are needed only to guarantee the gauge invariance of 
the matrix element. For physical applications one should take into account 
only the gauge invariant part of the diagram in Fig.~\ref{fig:X-em-decay}-a. 

\begin{figure}[htbp]
\vspace*{0.5cm}
\begin{center}
\includegraphics[scale=0.35]{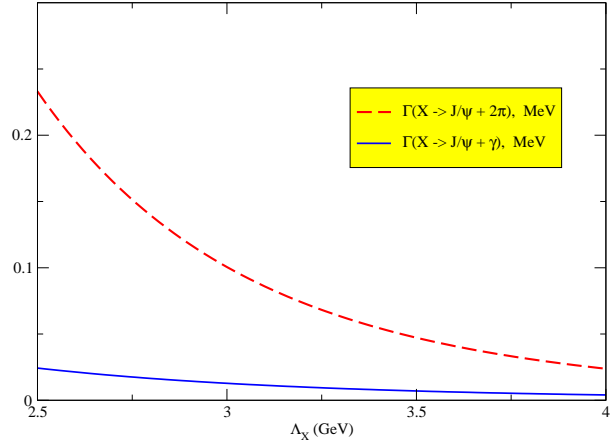}
\end{center}
\caption{
\label{fig:X-em-width}
The dependence of the decay widths 
$\Gamma(X_l\to J/\psi+\gamma)$ 
and $\Gamma(X_l\to J/\psi+2\pi)$ on the size parameter $\Lambda_X$.
}
\end{figure}

It is convenient to present the decay width via helicity or multipole
amplitudes. 
One has 
\begin{equation}
\Gamma(X\to J/\psi+\gamma) =\frac{1}{12\pi}\,
 \frac{|{\bf q_2}|}{m_X^2}\,\Big( |H_L|^2 +|H_{T}|^{2}\Big)
= \frac{1}{12\pi}\,\frac{|{\bf q_2}|}{m_X^2}\,\Big( |A_{E1}|^2 +|A_{M2}|^{2}\Big)
\label{eq:width}
\end{equation}
where the helicity amplitudes $H_L$ and $H_T$ are expressed in terms
of the Lorentz amplitudes as
\begin{eqnarray}
H_L &=& im_Xm_{J/\psi}|{\bf q_2}|
     \Big[ W_1 + \frac{m_X}{m_{J/\psi}^2}|{\bf  q_2}|\,W_3-W_4\Big]\,,
\nonumber\\
H_T &=& -im^2_{J/\psi}|{\bf q_2}|
      \Big[W_1+\frac{m_X}{m_{J/\psi}^2}|{\bf q_2}|\,W_2
        -\Big(1+\frac{m_X|{\bf q_2}|}{m^2_{J/\psi}}\Big)\,W_4\Big]\,,
\nonumber\\
&& |{\bf q_2}|=\frac{m_X^2-m^2_{J/\psi}}{2m_X}\,.
 \end{eqnarray}
Proceeding in a such way one will get the dependence of the decay widths
$\Gamma(X_l\to J/\psi+\gamma)$ and $\Gamma(X_l\to J/\psi+2\pi)$ plotted in 
Fig.~\ref{fig:X-em-width}. 

Note that the radiative decay width 
for $X_h=-X_u\sin\theta+X_d\cos\theta$ is almost an order of magnitude less 
than for $X_l=X_u\cos\theta+X_d\sin\theta$. 
If one takes $\Lambda_X\in (3,4)$~GeV with the middle point $\Lambda_X=3.5$~GeV
then the ratio of the widths is equal to

\begin{equation}
\frac{\Gamma(X_l\to J/\psi+\gamma)}
     {\Gamma(X_l\to J/\psi+2\pi)}\Big|_{\rm theor} = 0.15\pm 0.03 
\label{eq:theory}
\end{equation}
which fits very well the experimental data from the BELLE collaboration
written down in their Eq.~(\ref{eq:expt}).

\begin{equation}
\frac{\Gamma(X\to J/\psi+\gamma)}
     {\Gamma(X\to J/\psi+2\pi)}     = \left\{
 \begin{array}{rl}
0.14 \pm 0.05   & \mbox{ BELLE\,\cite{Abe:2005ix} } \\[2ex]
0.22 \pm 0.06   & \mbox{ BABAR\,\cite{Klempt:2007cp} } 
\end{array}\right.
\label{eq:expt}
\end{equation}

\begin{figure}[htbp]
\begin{center}
\includegraphics[scale=0.40]{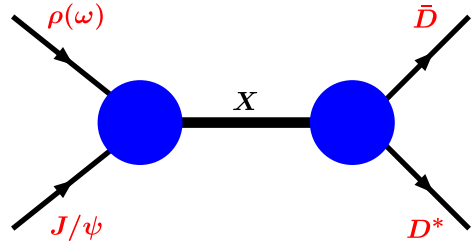}
\end{center} 
\caption{
\label{fig:Jpsi-X-diss}
Diagram describing the X-resonance contribution
to the $J/\psi$-dissociation process.}
\end{figure}

The last topic which we would like to discuss is the impact
of the intermediate X-resonance on the value of the $J/\psi$-dissociation
cross section, see \cite{Barnes:2003vt}-\cite{Blaschke:2009uh}. 
The relevant s-channel diagram is shown in  Fig.~\ref{fig:Jpsi-X-diss}.  

\begin{figure}[htbp]
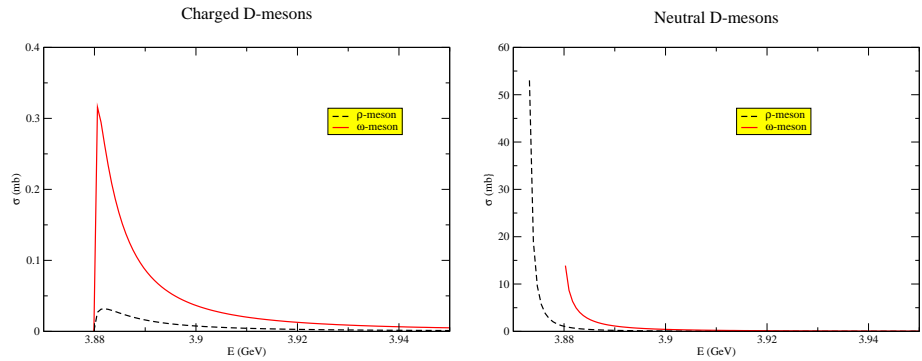

\begin{center}
\begin{tabular}{lr}
\includegraphics[scale=0.25]{J_dis_charged.eps} &
\includegraphics[scale=0.25]{J_dis_neutral.eps}
\end{tabular}
\end{center} 
\caption{
\label{fig:Jpsi-X-diss-width}
The cross sections of the processes
$J/\psi+v^0\to X \to D + D^{\ast}$. Charged D-mesons--
left panel, neutral D-mesons--right panel.}
\end{figure}

We take $\Gamma_X=1$~MeV in the Breit-Wigner propagator and
set $\Lambda_X=3.5$~GeV when calculating the matrix elements.
We plot the behavior of the relevant cross sections in 
Fig.~\ref{fig:Jpsi-X-diss-width}.
One can see that in the case of charged D-mesons (left panel
in Fig.~\ref{fig:Jpsi-X-diss-width})  the maximum value of the cross section
is about 0.32~mb at  $E=3.88$~GeV. This result should be 
compared with the result of the cross section
$\sigma(J/\psi+\pi\to D + \bar D^{\ast})\approx 0.9$~mb at 
$E=4.0$~GeV, see, \cite{Ivanov:2003ge} and the result of the cross section
$\sigma(J/\psi+\rho\to D + \bar D^{\ast})\approx 2.9$~mb at 
$E=3.9$~GeV, see, \cite{Barnes:2003vt}. 
Thus the X-resonance gives a sizable contribution
to the $J/\psi$-dissociation cross section. 
It would be interesting to do a complete analysis of the $J/\psi$
dissociation cross section in view of our new results on the
s-channel contribution of the X(3872) tetraquark state. 


\clearpage
\section{Summary}

\begin{itemize}

\item We have presented a refined covariant quark model which
includes infrared confinement of quarks.

\item We have calculated the transition form factors 
of the heavy $B_{s}-$meson to light pseudoscalar and vector mesons,
which are needed as ingredients for the calculation of the semileptonic,
nonleptonic, and rare decays. Our form factor
results hold in the full kinematical range of momentum transfer. 

\item We have made use of the calculated form factors to
calculate the nonleptonic decays 
$B_{s} \to D_{s} \bar D_{s}, ...$
and $B_{s} \to J/\psi\phi$, 
which have been widely discussed recently in the
context of $B_s-\bar B_s$--mixing and CP violation.  

\item
We have applied our approach to baryon physics by using the same values
of the constituent quark masses and infrared cutoff as in meson sector.

\item
We have calculated the nucleon magnetic moments and charge radii
and also electromagnetic form factors at low energies.

\item  The properties of the $X(3872)$ as a tetraquark
have been studied in the framework of a covariant quark model
with infrared confinement. 

\item The matrix elements of the off-shell transitions
$X\to J/\psi +\rho(\omega)$ and 
$X\to D + \bar D^\ast$ were calculated. 

\item The obtained results were then used to evaluate the widths 
of the experimentally observed decays
$X\to J/\psi+2\pi(3\pi)$ and 
$X\to D^0+\bar D^0 + \pi^0$. 

\item The possible impact of the $X(3872)$   
on the $J/\psi$-dissociation process was disscussed.

\item 
We have calculated  the matrix element of the transition
$X\to \gamma + J/\psi $ and have shown its gauge 
invariance. We have evaluated the $X\to \gamma + J/\psi $
decay width and the polarization
of the  $J/\psi$ in the decay.

\item The comparison with available experimental data
allows one to conclude that the $X(3872)$
can be a tetraquark state.
\end{itemize}

\end{document}